\documentclass[sigconf]{acmart}

\usepackage{amsfonts}
\usepackage{textcomp}
\usepackage{xcolor}
\usepackage{multirow}
\usepackage{enumitem}
\usepackage{listings}
\usepackage{tikz}
\usepackage[caption=false,font=footnotesize]{subfig}
\usepackage{stfloats}
\usepackage{makecell}
\usepackage{pifont}
\usepackage{acmart-taps}

\usepackage[ruled]{algorithm2e}

\usepackage[nogroupskip, numberedsection, acronym]{glossaries}
\makenoidxglossaries

\newacronym{mpc}{MPC}{Multi-Party Computation}
\newacronym{put}{PUT}{Public Transaction}
\newacronym{prt}{PRT}{Private Transaction}
\newacronym{mpt}{MPT}{Multi-Party Transaction}
\newacronym{mpp}{MPP}{Multi-Party Program}
\newacronym{zkp}{ZKP}{Zero-Knowledge Proof}
\newacronym{he}{HE}{Homomorphic Encryption}
\newacronym{tee}{TEE}{Trusted Execution Environment}
\newacronym{dpc}{DPC}{Decentralized Private Computation}
\newacronym{pop}{PoP}{Proof of Publication}
\newacronym{cca2}{IND-CCA2}{Indistinguishability under Adaptive CCA}
\newacronym{cma}{EUF-CMA}{Existential Unforgeability under CMA}
\newacronym{ttp}{TTP}{Trusted Third Party}

\usepackage{xcolor}
\definecolor{pink}{RGB}{255,160,122}
\definecolor{main}{RGB}{221, 157, 118} 

\newcommand{\codename}[0]{\textsc{Cloak}\xspace}
\newcommand{\myparagraph}[1]{\noindent{\textbf{#1.}}}

\newcommand{\etal}{\hbox{\emph{et al.}}\xspace}

\newcommand{\eg}{\hbox{\emph{e.g.}}\xspace}

\newcommand{\ie}{\hbox{\emph{i.e.}}\xspace}

\newcommand{\etc}{\hbox{\emph{etc.}}\xspace}

\newcommand{\longsquiggly}{\xymatrix{{}\ar@{~>}[r]&{}}}

\newcommand{\code}[1]{{\texttt{\small #1}}}

\usepackage[english]{babel}
\usepackage{amsthm}
\theoremstyle{theorem}
\newtheorem{informal-theorem}{Theorem}

\newtheorem{theorem}{Theorem}

\newtheorem{definition}{Definition}

\usepackage{listings, xcolor}

\definecolor{verylightgray}{rgb}{.97,.97,.97}

\lstdefinelanguage{Solidity}{
	keywords=[1]{anonymous, assembly, assert, balance, break, call, callcode, case, catch, class, constant, continue, constructor, contract, debugger, default, delegatecall, delete, do, else, emit, event, experimental, export, external, false, finally, for, function, gas, if, implements, import, in, indexed, instanceof, interface, internal, is, length, library, log0, log1, log2, log3, log4, memory, modifier, new, payable, pragma, private, protected, public, pure, push, require, return, returns, revert, selfdestruct, send, solidity, storage, struct, suicide, super, switch, then, this, throw, transfer, true, try, typeof, using, value, view, while, with, addmod, ecrecover, keccak256, mulmod, ripemd160, sha256, sha3}, 
	keywordstyle=[1]\bfseries,
	keywords=[2]{address, bool, byte, bytes, bytes1, bytes2, bytes3, bytes4, bytes5, bytes6, bytes7, bytes8, bytes9, bytes10, bytes11, bytes12, bytes13, bytes14, bytes15, bytes16, bytes17, bytes18, bytes19, bytes20, bytes21, bytes22, bytes23, bytes24, bytes25, bytes26, bytes27, bytes28, bytes29, bytes30, bytes31, bytes32, enum, int, int8, int16, int24, int32, int40, int48, int56, int64, int72, int80, int88, int96, int104, int112, int120, int128, int136, int144, int152, int160, int168, int176, int184, int192, int200, int208, int216, int224, int232, int240, int248, int256, mapping, string, uint, uint8, uint16, uint24, uint32, uint40, uint48, uint56, uint64, uint72, uint80, uint88, uint96, uint104, uint112, uint120, uint128, uint136, uint144, uint152, uint160, uint168, uint176, uint184, uint192, uint200, uint208, uint216, uint224, uint232, uint240, uint248, uint256, var, void, ether, finney, szabo, wei, days, hours, minutes, seconds, weeks, years},	
	keywordstyle=[2]\color{teal}\bfseries,
	keywords=[3]{block, blockhash, coinbase, difficulty, gaslimit, number, timestamp, msg, data, gas, sender, sig, value, now, tx, gasprice, origin},	
	keywordstyle=[3]\color{violet}\bfseries,
	keywords=[4]{k, @k, p, @p, x, @x, @me, @tee, @all, @winner}, 
	keywordstyle=[4]\color{main}\bfseries,
	keywords=[5]{reveal},	
	keywordstyle=[5]\color{main}\bfseries,
	identifierstyle=\color{black},
	sensitive=false,
	comment=[l]{//},
	morecomment=[s]{/*}{*/},
	commentstyle=\color{gray}\ttfamily,
	stringstyle=\color{red}\ttfamily,
	morestring=[b]',
	morestring=[b]"
}

\colorlet{punct}{red!60!black}
\definecolor{background}{HTML}{EEEEEE}
\definecolor{delim}{RGB}{20,105,176}
\colorlet{numb}{magenta!60!black}

\lstdefinelanguage{json}{
    basicstyle=\footnotesize\ttfamily,
    numbers=left,
    numberstyle=\footnotesize,
    stepnumber=1,
    numbersep=9pt,
    showstringspaces=false,
    captionpos=b,
	mathescape=true,
	tabsize=2,
	showtabs=false,
    literate=
     *{0}{{{\color{numb}0}}}{1}
      {1}{{{\color{numb}1}}}{1}
      {2}{{{\color{numb}2}}}{1}
      {3}{{{\color{numb}3}}}{1}
      {4}{{{\color{numb}4}}}{1}
      {5}{{{\color{numb}5}}}{1}
      {6}{{{\color{numb}6}}}{1}
      {7}{{{\color{numb}7}}}{1}
      {8}{{{\color{numb}8}}}{1}
      {9}{{{\color{numb}9}}}{1}
      {:}{{{\color{punct}{:}}}}{1}
      {,}{{{\color{punct}{,}}}}{1}
      {\{}{{{\color{delim}{\{}}}}{1}
      {\}}{{{\color{delim}{\}}}}}{1}
      {[}{{{\color{delim}{[}}}}{1}
      {]}{{{\color{delim}{]}}}}{1},
}
\AtBeginDocument{%
  }

\copyrightyear{2022}
\acmYear{2022}
\setcopyright{acmcopyright}\acmConference[ACSAC '22]{Annual Computer Security Applications Conference}{December 5--9, 2022}{Austin, TX, USA}
\acmBooktitle{Annual Computer Security Applications Conference (ACSAC '22), December 5--9, 2022, Austin, TX, USA}
\acmPrice{15.00}
\acmDOI{10.1145/3564625.3567995}
\acmISBN{978-1-4503-9759-9/22/12}

\begin{document}

\title{\textsc{Cloak}: Transitioning States on Legacy Blockchains Using Secure and Publicly Verifiable Off-Chain \acrlong{mpc}}


\author{Qian Ren}
\email{qianren1024@gmail.com}
\affiliation{%
  \institution{SSC Holding Company Ltd.}
  \city{Chengmai}
  \country{China}
}
\affiliation{%
  \institution{Oxford-Hainan Blockchain Research Institute}
  \city{Chengmai}
  \country{China}
}

\author{Yingjun Wu}
\email{yingjun@oxhainan.org}
\affiliation{%
  \institution{SSC Holding Company Ltd.}
  \city{Chengmai}
  \country{China}
}
\affiliation{%
  \institution{Oxford-Hainan Blockchain Research Institute}
  \city{Chengmai}
  \country{China}
}

\author{Han Liu}
\email{liuhan0518@163.com}
\affiliation{%
  \institution{Oxford-Hainan Blockchain Research Institute}
  \city{Chengmai}
  \country{China}
}
\affiliation{%
  \institution{Tsinghua University}
  \city{Beijing}
  \country{China}
}

\author{Yue Li}
\email{liyue@oxhainan.org}
\affiliation{%
  \institution{Oxford-Hainan Blockchain Research Institute}
  \city{Chengmai}
  \country{China}
}

\author{Anne Victor}
\email{anne.victor@outlook.com}
\affiliation{%
  \institution{SSC Holding Company Ltd.}
  \city{Chengmai}
  \country{China}
}
\affiliation{%
  \institution{Oxford-Hainan Blockchain Research Institute}
  \city{Chengmai}
  \country{China}
}

\author{Hong Lei}
\email{leihong@oxhainan.org}
\affiliation{%
  \institution{Hainan University}
  \city{Haikou}
  \country{China}
}
\affiliation{%
  \institution{Oxford-Hainan Blockchain Research Institute}
  \city{Chengmai}
  \country{China}
}

\author{Lei Wang}
\email{wanglei@cs.sjtu.edu.cn}
\affiliation{%
  \institution{Shanghai Jiao Tong University}
  \city{Shanghai}
  \country{China}
}

\author{Bangdao Chen}
\email{bangdao@oxhainan.org}
\affiliation{%
  \institution{SSC Holding Company Ltd.}
  \city{Chengmai}
  \country{China}
}
\affiliation{%
  \institution{Oxford-Hainan Blockchain Research Institute}
  \city{Chengmai}
  \country{China}
}

%
\renewcommand{\shortauthors}{Ren, et al.}

%
\begin{abstract}
    In recent years, the confidentiality of smart contracts has become a fundamental requirement for practical applications. While many efforts have been made to develop architectural capabilities for enforcing confidential smart contracts, a few works arise to extend confidential smart contracts to \acrfull{mpc}, \ie, multiple parties jointly evaluate a transaction off-chain and commit the outputs on-chain without revealing their secret inputs/outputs to each other. However, existing solutions lack public verifiability and require $O(n)$ transactions to enable negotiation or resist adversaries, thus suffering from inefficiency and compromised security.

In this paper, we propose \codename, a framework for enabling \acrfull{mpt} on existing blockchains. An \acrshort{mpt} refers to transitioning blockchain states by an \textit{publicly verifiable} off-chain \acrshort{mpc}.
We identify and handle the challenges of securing \acrshort{mpt} by harmonizing \acrshort{tee} and blockchain. Consequently, \codename secures the off-chain nondeterministic negotiation process (a party joins an \acrshort{mpt} without knowing identities or the total number of parties until the \acrshort{mpt} proposal settles), achieves public verifiability (the public can validate that the \acrshort{mpt} correctly handles the secret inputs/outputs from multiple parties and reads/writes states on-chain), and resists Byzantine adversaries. According to our proof, \codename achieves better security with only 2 transactions, superior to previous works that achieve compromised security at $O(n)$ transactions cost. By evaluating examples and real-world \acrshort{mpt}s, the gas cost of \codename reduces by 32.4\% on average. 
\end{abstract}

%
%






\maketitle


\vspace{-0.15cm}
\section{Introduction} \label{sec:introduction}
    \begin{table*}[!htbp]
    \setlength{\belowcaptionskip}{-0.5cm}
    \centering
    \footnotesize
    \caption{\textbf{Comparison of \codename with related works.} \small{Here, \ding{108}, \ding{119}, \ding{109}, \ding{53} denotes full, partial, not matched and not related, respectively. ``Adversary Model'' denotes how many entities' misbehavior are considered, where an executor denotes a server hosting \acrshort{tee}. ``min(\#TX)'' denotes how many transactions are required by the approach. ``Public Verifiability'' denotes all elements are committed on-chain and state transition can be validated, where $x$ denotes transaction parameter, $s, s'$ denotes contract old and new states respectively, $f$ denotes target function, $r$ denotes return value, and $\mathcal{P}$ denotes privacy policy that includes party-input bindings, \etc ``Financial Fairness'' denotes that honest parties never lose their collateral without obtaining outputs.}}
    \label{tab:comparison}
    \setlength{\tabcolsep}{1.7mm}{
    \begin{tabular}{lcccccccccccccccc}
        \toprule
        \multirow{2}{*}{\textbf{Approach}} & \multicolumn{2}{c}{\textbf{Adversary Model}} && \multirow{2}{*}{\textbf{\makecell[c]{Chain\\Agnostic}}} & \multirow{2}{*}{\textbf{min(\#TX)}} & \multirow{2}{*}{\textbf{Confidentiality}} & \multirow{2}{*}{\textbf{\makecell[c]{Nondeterministic\\Negotiation}}} & \multicolumn{6}{c}{\textbf{Public Verifiability}} && \multirow{2}{*}{\textbf{\makecell[c]{Financial\\Fairness}}} \\
        \cmidrule{2-3}\cmidrule{9-14} & \#Parties & \#Executors && & & & & $x$ & $s$ & $f$ & $r$ & $s'$ & $\mathcal{P}$ && \\
        \midrule
        Ethereum~\cite{wood2014ethereum} & $1^*$ & \ding{53} && \ding{53} &$O(1)$ & \ding{53} & \ding{53} & \ding{108} & \ding{108} & \ding{108}  & \ding{108} & \ding{108} & \ding{108} && \ding{53}  \\
        Ekiden~\cite{Ekiden:2019}& $1^*$ & $m^*-1^1$ && \ding{108} & $O(1)$ & \ding{108} & \ding{53} & \ding{109}$^2$ & \ding{108} & \ding{108}  & \ding{109}$^2$ & \ding{108}  & \ding{108} && \ding{53} \\
        Confide~\cite{CONFIDE:SIGMOD20}& $1^*$ & $\lfloor m^*/3 \rfloor^3$ && \ding{109} & $O(1)$ & \ding{108} & \ding{53} & \ding{108} & \ding{108} & \ding{108} & \ding{108} & \ding{108} & \ding{108} && \ding{53} \\
        Hawk~\cite{Hawk:SP2016}& $n^*$ & \ding{53} && \ding{108} & $O(n)$ & \ding{119}$^4$ & \ding{109} & \ding{108} & \ding{109} & \ding{108} & \ding{108} & \ding{109} & \ding{109} && \ding{108} \\
        ZEXE~\cite{ZEXE:SP20} & $n^*$ & $1^*$ && \ding{109} & $O(1)$ & \ding{119} & \ding{109} & \ding{108} & \ding{108} & \ding{108} & \ding{108} & \ding{108} & \ding{109} && \ding{53} \\
        Fastkitten~\cite{FastKitten'19}& \multicolumn{2}{c}{$(n^*+1^*)-1$} && \ding{109} & $O(n)$ & \ding{119} & \ding{109} & \ding{109} & \ding{109} & \ding{109} & \ding{108} & \ding{109} & \ding{109} && \ding{108} \\
        LucidiTEE~\cite{Sinha2020LucidiTEEAT}& $n^*$ & $m^*-1$ && \ding{108} & $O(n)$ & \ding{108} & \ding{108} & \ding{108} & \ding{119}$^5$ & \ding{108} & \ding{108} & \ding{119}$^5$ & \ding{119}$^5$ && \ding{53} \\
        \textbf{\codename} & \multicolumn{2}{c}{$(n^*+1^*)-1^6$} && \ding{108} & $O(1)$ & \ding{108} & \ding{108} & \ding{108} & \ding{108} & \ding{108} & \ding{108} & \ding{108} & \ding{108} && \ding{108} \\
        \bottomrule
    \end{tabular}
    \\
    {${}^1$ The $^*$ denotes the total number of specific kinds of entities assumed in the system, \eg, $1^*$ denotes the unique party/executor, $n^*$ denotes all $n$ parties, and $m^*$ denotes all executors in the system.
    ${}^2$ Transaction parameters $x$ (resp. return values $r$) in Ekiden are received (resp. delivered) off-chain while not committed on-chain.
    ${}^3$ We assume Confide's undeclared consensus is BFT. 
    ${}^4$ The manager is expected not to leak parties' private data. 
    ${}^5$ Fastkitten does not commit the inputs and states of \acrshort{mpt}. 
    ${}^6$ LucidiTEE does not consider verifying the state transition with policy on-chain.}
    } 
    \vspace{-0.4cm} 
\end{table*}

With the rapid development of blockchains, privacy issues have become 
one of the top concerns for smart contracts. 
Unfortunately, despite the importance of smart contract privacy, most existing blockchains are 
designed \emph{without privacy} by nature~\cite{nakamoto2008bitcoin, wood2014ethereum}. For example, miners of Ethereum verify transactions in a block 
by re-executing them with the exact input and states. Consequently, the transaction data have to be shared within the entire network.


\myparagraph{\emph{Confidential smart contract with \acrshort{mpc}}} 
To address the aforementioned problem, researchers have proposed 
various \emph{confidential smart contract} solutions, \ie, keeping transaction inputs and contract states secret from  
non-participants.
In parallel, a few works expand the transaction of confidential smart contracts to \acrfull{mpc}s, which means allowing multiple parties jointly evaluate a transaction off-chain and commit the outputs on-chain without revealing their secret inputs/outputs to each other. These works fall into two categories. The first adopts cryptographic \acrshort{mpc} primitives (based on secret sharing~\cite{Sanchez2018MPCContract}, \acrshort{he}~\cite{SamuelSP2022ZeeStar}, and \acrshort{zkp}~\cite{Hawk:SP2016, ZEXE:SP20}, \etc) to let multiple parties jointly evaluate a transaction off-chain and optionally record or partially prove the evaluation on-chain. The other category adopts \acrshort{tee} to collect sealed inputs from different parties, reveal the inputs and evaluate a program inside enclaves to obtain the outputs~\cite{FastKitten'19, liaoAsiaCCS2022speedster}. While both categories achieve confidentiality of \acrshort{mpc}, few of them achieve public verifiability. Qian \etal~\cite{ren2021CloakDemo} call the need for publicly verifiable \acrshort{mpc} transaction in reason that the transaction should prove to non-participants of its \acrshort{mpc}, especially regulators or miners, to let them trust the state transition the transaction caused. Qian \etal~\cite{ren2021CloakDemo} furthermore firstly define a problem \acrfull{mpt}, which refers to multiple parties jointly evaluating a transaction off-chain based on publicly verifiable \acrshort{mpc} to transition states on-chain, while keeping each secret input/output confidential to its corresponding party. 

\myparagraph{\emph{Limitations and Challenges}} In this paper, we aim to support \acrshort{mpt}-enabled confidential smart contracts on legacy blockchains, which poses several challenges that existing efforts fail to handle.

\textbf{C1: Nondeterministic negotiation with minimal cost}. 
In the real world, users should be allowed to join an \acrshort{mpt} without knowing other parties' information prior, \eg, bidders can independently decide to join an auction, as the number and identities of all bidders are settled only when the bidding phase closes. We call this nondeterministic negotiation. However, practically secure nondeterministic negotiation is nontrivial. Previous approaches either assume that protocols start with pre-known settings~\cite{Hawk:SP2016, Bhavani'17-FariMPC, FastKitten'19} (program, parties, time duration, \etc) to bypass the challenge, or require each party to send transactions on-chain thereby causing $O(n)$ transactions~\cite{Sinha2020LucidiTEEAT}), or assume parties negotiate by P2P communications or assistance of a semi-honest \acrfull{ttp}~\cite{tss-lib}, thus vulnerable to Byzantine adversary. Consequently, securing off-chain negotiation under Byzantine adversary at the cost of $O(1)$ transactions is still a challenge.

\textbf{C2: Publicly verifiable \acrshort{mpc}-based state transition}. 
To transition blockchain states by off-chain \acrshort{mpc}s~\cite{Sanchez2018MPCContract, FastKitten'19, Sinha2020LucidiTEEAT}, Qian \etal~\cite{ren2021CloakDemo} stress that the public, including miners, should also verify the state transition the \acrshort{mpc} caused without trusting any parties or executors of the \acrshort{mpc}, and identifies the problem as a new problem \acrshort{mpt}. However, Qian \etal~\cite{ren2021CloakDemo} fails to present an corresponding capable and secure protocol. Existing cryptographic solutions for \acrshort{mpc} under malicious adversaries perform well on achieving confidentiality, but cannot achieve the public verifiability required by \acrshort{mpt}. Specifically,~\cite{FastKitten'19, Sinha2020LucidiTEEAT, Bhavani'17-FariMPC, kanjalkarEuroSPW21} sporadically record part information of an off-chain \acrshort{mpc} evaluation (inputs, outputs, states, \etc) on the blockchain, failing to uniquely identify the evaluation, not to mention prove it. Moreover, for miners/regulators who neither are nor trust any \acrshort{mpc} participants, the participants cannot only record or multi-sign messages to convince miners/regulators that the \acrshort{mpc}-caused state transition holds authenticity and correctness. Consequently, it is still a challenge to construct a general and succinct $proof$ to achieve \acrshort{mpt}s.

\textbf{C3: Byzantine adversary resistance with minimal cost}. To punish off-chain misbehaviors like aborting protocols, previous work~\cite{MPCpenalties'16, MPCBitcoin'16, PokerBitcoin'15, FastKitten'19} involve fine-tuned challenge-response mechanisms. These mechanisms require all parties to independently deposit collateral on-chain before the protocol starts, thus leading to at least $O(n)$ transactions, which is impractical for scalability.

\myparagraph{\emph{Our work}} In this paper, we propose a novel \acrshort{mpt}-enabled confidential smart contract framework, \codename, by harmonizing the blockchain with a unique \acrshort{tee}-enabled executor. Furthermore, we design and prove a currently most secure and practical protocol for serving \acrshort{mpt}, under the assumption that a Byzantine adversary can arbitrarily corrupt parties or the executor but cannot break the integrity of the \acrshort{tee} itself.

\myparagraph{\emph{Contributions}} 
Our main contributions are as follows: 

\begin{itemize}[leftmargin=3mm, parsep=0mm, topsep=0mm, partopsep=0mm]
    \item We propose a novel confidential smart contract framework, which can transition the state of existing blockchains by transactions based on publicly verifiable \acrshort{mpc}, \ie, \acrshort{mpt}.
    \item With Byzantine adversary assumed, we design a protocol to achieve trusted off-chain nondeterministic negotiation (against \textbf{C1}), public verifiability (against \textbf{C2}), and financial fairness (against \textbf{C3}) for \acrshort{mpt}, at the cost of only 2 transactions. 
    \item We formally define and prove the security properties that \codename achieved in a Byzantine adversary model.
    \item We have applied \codename in several real-world scenarios. \codename achieves \acrshort{mpt} with both lower gas costs and better performance.
\end{itemize}


\vspace{-0.15cm}
\section{Related work} \label{sec:related-work}
    In this section, we elaborate how \codename is distinct from current confidential smart contract solutions. Table~\ref{tab:comparison} shows the comparison between \codename and some representative solutions.

\myparagraph{TEE-enforced confidential smart contracts} 
Ekiden~\cite{Ekiden:2019, secondstate'20} reveals and executes smart contracts in SGX to conceal transaction parameters, return values, and contract states. CCF~\cite{ccf2019} supports any typescript or C++-based application in a TEE-based permissioned blockchain. Confide~\cite{CONFIDE:SIGMOD20} synchronizes a common public key between all SGX and runs EVM and WASM in SGX to support various contracts. 
CCF, Confide, and Ekiden integrate TEE into their own consensus pipeline, thus making them chain-specific. In contrast, \codename enables \acrshort{mpt} on an existing blockchain by deploying merely a contract facility and is thus chain-agnostic. 
For scalability, each transaction of~\cite{Ekiden:2019, secondstate'20, ccf2019, CONFIDE:SIGMOD20} is from a single sender and validated by all consensus nodes. Therefore, they do not consider negotiation, and serving \acrshort{mpt} with these solutions takes at least $O(n)$ transactions from parties.
For public verifiability, CCF and Confide ignore off-chain inputs/outputs. Ekiden considers off-chain inputs/outputs and can verify state transition on-chain. However, Ekiden does not commit off-chain data on-chain, therefore the transaction cannot be identified. Conversely, \codename considers and commits all necessary elements of \acrshort{mpt} evaluation on-chain, enabling it to identify the evaluation. 
For security, none of the above solutions can punish misbehaving transaction sender or executors. Instead, \codename secures off-chain inputs submission and outputs delivery, and achieves financial fairness.

\myparagraph{Cryptography-based smart contracts with \acrshort{mpc}}
A cryptography-based contract with~\acrshort{mpc} refers to multiple parties jointly evaluating a transaction based on cryptographic schemes.
In terms of scalability, \acrshort{mpc}-based approaches~\cite{MPCpenalties'16, MPCBitcoin'16, PokerBitcoin'15} allow $m$-round \acrshort{mpc} with penalties over Bitcoin but rely on claim-or-refund functionality, which necessitates complex and expensive transactions and collateral, thus making these solutions impractical. Hawk~\cite{Hawk:SP2016} also requires $O(n)$ transactions to punish misbehaved parties.
In terms of confidentiality, Hawk requires a credible manager to withhold information, thus achieving limited confidentiality. ZEXE~\cite{ZEXE:SP20} proves the satisfaction of predicates with \acrshort{zkp} without revealing party secrets to the public. However, to generate the proof for a predicate, a party must be privy to the predicate's secrets, thus violating inter-party confidentiality. Instead, \codename keeps parties' confidential data confidential from each other and the public.
In terms of negotiation, both Hawk and ZEXE start from pre-specified parties, assuming that parties have previously negotiated through off-chain P2P communications or a \acrshort{ttp}.
In terms of public verifiability, the manager of Hawk updates contract state with \acrshort{zkp} proof, making the off-chain multi-party process transparent and unverifiable to verifiers, \eg, miners. Public auditable \acrshort{mpc} (PA-\acrshort{mpc})~\cite{CarstenSCN'14PAMPC} achieves the publicly verifiable \acrshort{mpc}, allowing multiple parties jointly evaluate a program and prove it. Nevertheless, existing PA-\acrshort{mpc} primitives are not designed for committing data or proving state transitions, \eg, \acrshort{mpc}s expressed in Solidity that operate both on- and off-chain inputs/outputs. Moreover, they have flaws at inefficiency and weaker adversary model, and still fail in practically supporting nondeterministic negotiation or achieving financial fairness. Specifically \cite{CarstenSCN'14PAMPC, BerryACNS15} requires trusted setup or un-corrupted parties. \cite{FoteiniASIACRYPT20} is function-limited. \cite{OzdemirSecurity22} very recently achieves general-purpose PA-\acrshort{mpc} but only support circuit-compatible operations. None of above solutions are for confidential smart contracts or can punish adversaries.

\myparagraph{TEE-based smart contracts with \acrshort{mpc}} 
For confidentiality, Fastkitten~\cite{FastKitten'19} does not consider confidentiality. 
For negotiation, both Fastkitten and Speedster~\cite{liaoAsiaCCS2022speedster} assume parties are known prior. LucidiTEE~\cite{Sinha2020LucidiTEEAT} requires parties to independently send transactions on-chain to join a \acrshort{mpt}; therefore, it achieves nondeterministic negotiation but flaws at $O(n)$ transactions.  \codename enables negotiation with the cost of 1 transaction.  
For public verifiability, both FastKitten and Speedster records only the final outputs of m-round \acrshort{mpc} on the blockchain. LucidiTEE does not distinguish between states of different parties. \codename proves and uniquely identifies the evaluation on-chain. 
For security, while LucidiTEE and Speedster lack collateral systems to punish malicious parties, both Hawk and FastKitten adopt challenge-response protocols to punish malicious parties. FastKitten achieves financial fairness but requires each party to deposit collateral for each evaluation, suffering from $O(n)$ transaction cost. In contrast, \codename achieves financial fairness with the cost remaining $O(1)$.

To the best of our knowledge, \codename is the first to enable existing blockchains to transition states by \acrshort{mpt}s. \codename is also advanced in securing off-chain nondeterministic negotiation process and resisting Byzantine adversary by normally only 2 transactions.

\vspace{-0.15cm}
\section{\codename overview} \label{sec:overview}
    This section presents an overview of \codename, a novel \acrshort{mpt}-enabled confidential smart contract framework. We first model the architecture of \codename, and then introduce the workflow of \codename, as well as our adversary model. Finally, we introduce our security goals and design challenges.

\begin{figure}[!h]
  \setlength{\belowcaptionskip}{-0.6cm} 
  \centering
  \includegraphics[width=7.5cm]{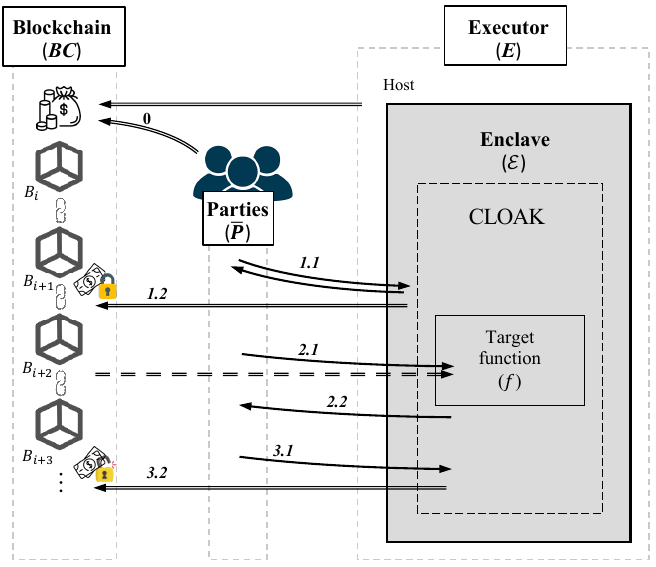}
  \caption{\textbf{Overall workflow of \codename framework}}
  \label{fig:framework}
\end{figure}

\vspace{-0.15cm}
\subsection{System model}
Conceptually, to support multiple parties evaluate \acrshort{mpt}. \codename adopts a hybrid architecture combining a blockchain and a unique executor with \acrshort{tee}. 
Figure~\ref{fig:framework} depicts the architecture of \codename and a workflow of the \codename protocol. There are mainly four entities: 

\myparagraph{Blockchain (\textbf{\textit{BC}})} A blockchain in \codename is a normal contract-enabled blockchain, \eg, Ethereum. It is responsible for maintaining the commitments of parameters, returns, states, and the \acrshort{mpc} of \acrshort{mpt} and validating the state transition.

\myparagraph{Parties (\={\textbf{\textit{P}}})} Parties are participants of \acrshort{mpt}. They interact with the \acrshort{tee} enclave to negotiate the \acrshort{mpt} setting, feed inputs and receive outputs. They also interact with the blockchain to monitor \acrshort{mpt} status and punish the misbehaving executor.

\myparagraph{Executor (\textit{\textbf{E}})} The executor is a server holding a \acrshort{tee}. It is responsible for instantiating the \acrshort{tee} enclave and relaying messages between parties, the blockchain and the enclave to proceed with the \codename protocol.

\myparagraph{Enclave ($\mathcal{E}$)} The enclave runs the \codename enclave program (Algorithm~\ref{alg:cloak-enclave}). It is responsible for receiving inputs from parties and the blockchain, evaluating the \acrshort{mpt} inside the enclave, and delivering the outputs to the blockchain and parties. 

\vspace{-0.15cm}
\subsection{\textbf{Adversary model}}

In our system, $n$ parties and a unique executor \textit{\textbf{E}} (who owns the \acrshort{tee} $\mathcal{E}$) follows the \codename protocol $\pi_{\codename}$ to enable \acrshort{mpt} on an existing blockchain. Our assumptions and threat model are as follows:

\myparagraph{\acrshort{tee}} We assume that the adversary cannot break the integrity and confidentiality of \acrshort{tee}. Although we instantiate the \acrshort{tee} as SGX, our design is \acrshort{tee}-agnostic. We stress that although recent research showed some attacks against \acrshort{tee}, the confidentiality and integrity guarantees of \acrshort{tee} are still trustworthy, making our assumption of \acrshort{tee} practical. We elaborate the rationality of this \acrshort{tee} assumption in Appendix \ref{sec:assumption-rationality}

\myparagraph{Blockchain} We assume the common prefix, chain quality, and chain growth of the blockchain are held so that the blockchain constantly processes and confirms new transactions and is always available. In particular, we assume that the blockchain supports Turing-complete smart contracts, \eg, Solidity, so that we can deploy a contract program (Algorithm \ref{alg:cloak-service}) on the blockchain to manage the life cycle of \acrshort{mpt}. Finally, while \codename is designed to be agnostic to the underlying consensus protocol, \textit{we assume the blockchain can construct a \acrfull{pop} of transactions for proving that a transaction has been confirmed on the blockchain, which is similarly assumed by~\cite{Cavallaro2019Tesseract, Ekiden:2019, FastKitten'19} for resisting Eclipse attacks}.

\myparagraph{Parties} Honest parties trust their own platforms but never trust other parties or \textit{\textbf{E}}. Honest parties trust the data accessed from blockchain and attested \acrshort{tee}. 

\myparagraph{Threat model} We assume a \textit{Byzantine adversary} is present in our system. The adversary can corrupt all but one subject among parties and \textit{\textbf{E}}. On compromised parties or \textit{\textbf{E}}, the adversary can behave arbitrarily, \eg, scheduling processes as well as reordering, delaying, or mutating messages but can never break the integrity and confidentiality of \acrshort{tee}. 

\vspace{-0.15cm}
\subsection{Design goals}
We aim to achieve the following five security properties. These properties are formally defined in Appendix~\ref{sec:security-definition}.

\myparagraph{Correctness} If an \acrshort{mpt} succeeds, its outputs are the correct outputs of the program, the target function, applied to the committed inputs.

\myparagraph{Confidentiality} Without any compromised \acrshort{tee}, \codename guarantees that both the inputs and outputs of \acrshort{mpt} are kept secret to their corresponding parties. 

\myparagraph{Public verifiability} The public with only on-chain data can verify that a state transition is correctly caused by an \acrshort{mpt}, and the \acrshort{mpt} is jointly evaluated using committed program, privacy policy, old states, parameters, causing committed return values and new states.

\myparagraph{Executor balance security} If the executor \textit{\textbf{E}} honestly behaves, it cannot lose money.

\myparagraph{Financial Fairness} Honest parties should never lose their collateral. Specifically, if at least one subject among both parties and \textit{\textbf{E}}, is honest, then either (i) the negotiation failed, and all parties stay financial neutral, or (ii) the protocol correctly evaluates the \acrshort{mpt} and all parties stay financial neutral, or (iii) all honest subjects among parties and \textit{\textbf{E}} know the protocol has aborted and stay financially neutral and at least one malicious subject among parties and \textit{\textbf{E}}, must be financially punished.

In addition to the foregoing goals of this paper, we also clarify what are not our goals here. The confidentiality broken by parties voluntarily revealing their own secrets to anyone else except \acrshort{tee} lies outside our consideration. Resisting the potential secret leakage caused by the \acrshort{mpc}'s target function is also not our goal.

\vspace{-0.15cm}
\subsection{Workflow}

As shown in Figure~\ref{fig:framework}, parties follow the \codename protocol to interact with the \textbf{\textit{BC}} and \textit{\textbf{E}} in order to send \acrshort{mpt}. Our protocol proceeds in four phases. In particular, while the setup phase (\textbf{\textit{0}}) occurs only once for the \textbf{\textit{E}} and each party, the remaining phases (\textbf{\textit{1-3}}) are repeated for each \acrshort{mpt}. We assume that the public key $pk_{\mathcal{E}}$ and address $adr_{\mathcal{E}}$ of enclave $\mathcal{E}$ are both previously registered on-chain. All parties have verified that $\mathcal{E}$ has been initialised correctly by inspecting its attestation report. Utilizing verified $pk_{\mathcal{E}}$, parties can establish secure channels with $\mathcal{E}$. In the following, as \textbf{\textit{E}} is responsible for relaying input and output messages of its $\mathcal{E}$, we illustrate the protocol in the same way that $\mathcal{E}$ directly communicates with \textbf{\textit{BC}} and \={\textbf{\textit{P}}}, rather than explicitly marking \textbf{\textit{E}}'s relaying behaviour each time. 

\begin{itemize}[leftmargin=3mm, parsep=0.5mm, topsep=0.5mm, partopsep=0.5mm]
    \item \textbf{(Global) Setup phase (\textbf{\textit{0}})}: \textbf{\textit{E}} and all parties independently deposit their coins to $\mathcal{E}$'s address $adr_{\mathcal{E}}$. In subsequent phases, the collateral required to join each \acrshort{mpt} is deducted from these coins. 
    
    \item \textbf{(\acrshort{mpt}) Negotiation phase (\textbf{\textit{1.1-1.2}})}: One of the parties sends an \acrshort{mpt} proposal to $\mathcal{E}$ without knowing the identities of other parties. Then, $\mathcal{E}$ starts a \textit{nondeterministic negotiation protocol} (\textbf{\textit{1.1}}). To begin, $\mathcal{E}$ generates an id for the proposal and broadcasts the signed proposal with the id to all parties. If any party is interested in or required by the proposal, it responds an acknowledgment to $\mathcal{E}$ with signed commitments of parameters. $\mathcal{E}$ continues to collect parties' acknowledgments until the negotiation phase ends or the collected acknowledgments satisfy the negotiation's settlement condition. Following that, $\mathcal{E}$ sends a $TX_p$ to \textbf{\textit{BC}} (\textbf{\textit{1.2}}). The $TX_p$ publishes the \acrshort{mpt} proposal along with all parties' identities and parameter commitments to indisputably announce the negotiation outcome on-chain. Additionally, $TX_p$ deducts collateral from \textbf{\textit{E}} and all parties for the \acrshort{mpt} in case any of them aborts the \acrshort{mpt} after the negotiation succeeds.
    
    \item \textbf{(\acrshort{mpt}) Execution phase (\textbf{\textit{2.1-2.2}})}: After $TX_p$ is confirmed on \textbf{\textit{BC}}, each party sends their signed inputs (\ie, parameters and old states) to $\mathcal{E}$. Upon receiving these inputs, $\mathcal{E}$ reads the blockchain view to ensure that $TX_p$ was indeed been confirmed on \textbf{\textit{BC}}. The confirmation of $TX_p$ indicates that parties' collateral has been successfully deducted
    and parameters committed (\textbf{\textit{2.1}}). Then, $\mathcal{E}$ verifies that if the collected parameters and old states match their on-chain commitments. If the verification succeeds, $\mathcal{E}$ evaluates the \acrshort{mpt} program to obtain outputs (\ie, return values and new states). Following that, $\mathcal{E}$ only delivers the ciphertext of outputs to parties off-chain (\textbf{\textit{2.2}}).
    
    \item \textbf{(\acrshort{mpt}) Distribution phase (\textbf{\textit{3.1-3.2}})}: After $\mathcal{E}$ collects all parties' receipts (\textbf{\textit{3.1}}), $\mathcal{E}$ sends a $TX_{com}$ to commit the outputs on \textbf{\textit{BC}} along with the encryption key of delivered output ciphertext  (\textbf{\textit{3.2}}). Thus, all parties accessing the key in $TX_{com}$ can decrypt the output ciphertext received in \textbf{\textit{2.2}}. 
\end{itemize}

\vspace{-0.15cm}
\subsection{\textbf{Design challenges and highlights}}
In this section, we highlight the challenges handled in the workflow and high-level ideas of their corresponding countermeasures.

\vspace{-0.15cm}
\subsubsection{Securing off-chain nondeterministic negotiation (against \textbf{C1})}
In a decentralized and open network, there are undoubtedly scenarios in which a party joins an \acrshort{mpt} unaware of the other parties, \eg, a public auction in which bidders self-select to participate without knowing others until the bidding process closes. The nondeterministic negotiation is for parties to negotiate a \acrshort{mpt} proposal without knowing others until the proposal is settled. An \acrshort{mpt} proposal can be exemplified as $p'\gets(C_f, C_{\mathcal{P}}, q, \bar{\bm P}, C_{x})$, where $C_*$ denotes the hash commitment of $*$. Therefore, the proposal $p'$ specifies which \acrshort{mpt} $f$ to evaluate, which policy $\mathcal{P}$ to enforce, which parties $\bar{\bm P}$ are required to participate and their corresponding inputs $x$, and how much collateral of misbehaving parties to punish.
Previous works~\cite{Hawk:SP2016, Bhavani'17-FariMPC, FastKitten'19} assume that \acrshort{mpt} proposal is known prior. Although~\cite{Sinha2020LucidiTEEAT} enables parties to autonomously bind inputs on-chain to join a specific \acrshort{mpt} proposal, it incurs a cost of $O(n)$ transactions. One may believe that we can require all parties to communicate with a \acrshort{tee} off-chain in order to negotiate with other parties without interacting with a blockchain. However, if we do not settle the negotiation on-chain, the blockchain will neither know when the \acrshort{mpt} begins (which is critical for timeout judgement) nor capable of freezing all collateral of parties and the executor before the evaluation. Consequently, the adversary can arbitrarily drops or delays off-chain data without being identified or penalised. 

In this paper, we propose an \textit{nondeterministic negotiation} subprotocol to support the nondeterministic participation of \acrshort{mpt} parties. The main idea is to allow a party to initiate a negotiation process by sending an \acrshort{mpt} proposal. After parties agree on the \acrshort{mpt} proposal, they can send their acknowledgements and parameter commitments to join the \acrshort{mpt}. When the negotiation is complete, the \acrshort{tee} attaches party identities and parameter commitments to the proposal to obtain a settled proposal. The settled proposal is then published on the blockchain by \acrshort{tee}. Thus, both parties and \acrshort{tee} proceed to the next phase based on the blockchain-confirmed proposal. The blockchain knowing when the \acrshort{mpt} begins is capable of judging whether the \acrshort{mpt} timeouts.

\vspace{-0.15cm}
\subsubsection{Achieving public verifiability of \acrshort{mpt} (against \textbf{C2})}
The challenge of achieving public verifiability of \acrshort{mpt} is constructing an interpretable $proof$ whose size is independent of $x, s, f, r, s'$ and the privacy policy $\mathcal{P}$. $\mathcal{P}$ denotes meta-transaction settings, \eg, party-input bindings~\cite{Sinha2020LucidiTEEAT}. 

To create a succinct and general $proof$, we use \acrshort{tee} to endorse the enforcement of \acrshort{mpt}. Let $\mathcal{E}$ denote \acrshort{tee}. $\mathcal{E}$ is expected to receive inputs $x, s$, run $f$, deliver outputs $r, s'$, generate $proof$, and enforce $\mathcal{P}$ throughout the process. Let $H(*)$ denote \aptLtoX[graphic=no,type=env]{$\texttt{hash}(*)$}{$\FuncSty{hash}(*)$}. When $\mathcal{E}$ successfully evaluates an \acrshort{mpt}, $\mathcal{E}$ sends a signed transaction $TX_{com}$ that includes a $proof\gets[H_{C_{\mathcal{P}}}, H_{C_f}, H_{C_s}, H_{C_x}, H_{C_{s'}}, H_{C_r}]$.
The signed $proof$ demonstrates $\mathcal{E}$'s endorsement of the state transition from $s$ to $s'$ caused by \acrshort{mpt}. Thus, if $H_{C_{\mathcal{P}}}, H_{C_f}$, and $H_{C_s}$ in $TX_{com}$ match their previously registered commitments on-chain, the blockchain then accepts the state transition. 

\vspace{-0.15cm}
\subsubsection{Resisting adversary with minimal transactions (against \textbf{C3})}
The interaction of an \acrshort{tee} with the environment is controlled by the \textit{\textbf{E}}. As a result, a malicious \textit{\textbf{E}} can stop the \acrshort{tee} from running or present Eclipse Attacks~\cite{Ekiden:2019} during the protocol. Malicious parties can also abort at any point during the protocol to launch a DoS attack. \cite{FastKitten'19} allows parties to punish the aborted \textit{\textbf{E}} after a certain amount of time has passed. Because the system relies on Bitcoin-specific time-delay transactions, it cannot be used on other platforms. \cite{FastKitten'19} also uses an enhanced \emph{challenge-response} subprotocol to distinguish between the malicious \textit{\textbf{E}} dropping party inputs and malicious parties failing to submission inputs. However, all those works on defending against the foregoing attacks, such as~\cite{Hawk:SP2016, MPCpenalties'16, FastKitten'19}, require both parties and the \textit{\textbf{E}} to deposit collateral at the start of the protocol. Even if all parties and the \textit{\textbf{E}} are honest, these works result in $O(n)$ transactions for each \acrshort{mpt}, which is expensive and inefficient. 

In this paper, we adopt a \textit{challenge-response} subprotocol similar to~\cite{FastKitten'19} to identify adversary in the input submission phase. The idea behind both subprotocols is that the protocol penalise \textit{\textbf{E}} by default unless \textit{\textbf{E}} can show the \acrshort{tee} that it has publicly challenged parties on-chain but received no reply. These two subprotocols require $O(m)$ transactions when $m$ malicious parties present. Furthermore, we design a \textit{one-deposit-multiple-transact} method. The method requires only two constant transactions when all subjects behave honestly. Specifically, \textit{\textbf{E}} and parties globally deposit coins as account balances to an address managed by \acrshort{tee}. Before evaluating an \acrshort{mpt}, \acrshort{tee} only deducts \acrshort{mpt}-specific collateral from \acrshort{mpt}-involved party accounts by a $TX_p$. If the \acrshort{mpt} succeeds, the \acrshort{tee} refunds the frozen \acrshort{mpt}-specific collateral to parties via $TX_{com}$. As a result, each party depositing coins once can join (sequentially or concurrently) numerous \acrshort{mpt}, as long as the total amount of deducted \acrshort{mpt}-specific collateral does not exceed the amount of coins deposited by the party. Finally, because \emph{challenge-response} subprotocols are rarely executed due to the high financial cost of adversary, we achieve $O(1)$ transactions per \acrshort{mpt} in normal cases.

\vspace{-0.15cm}
\section{\codename protocol} \label{sec:cloak-deploy}
    
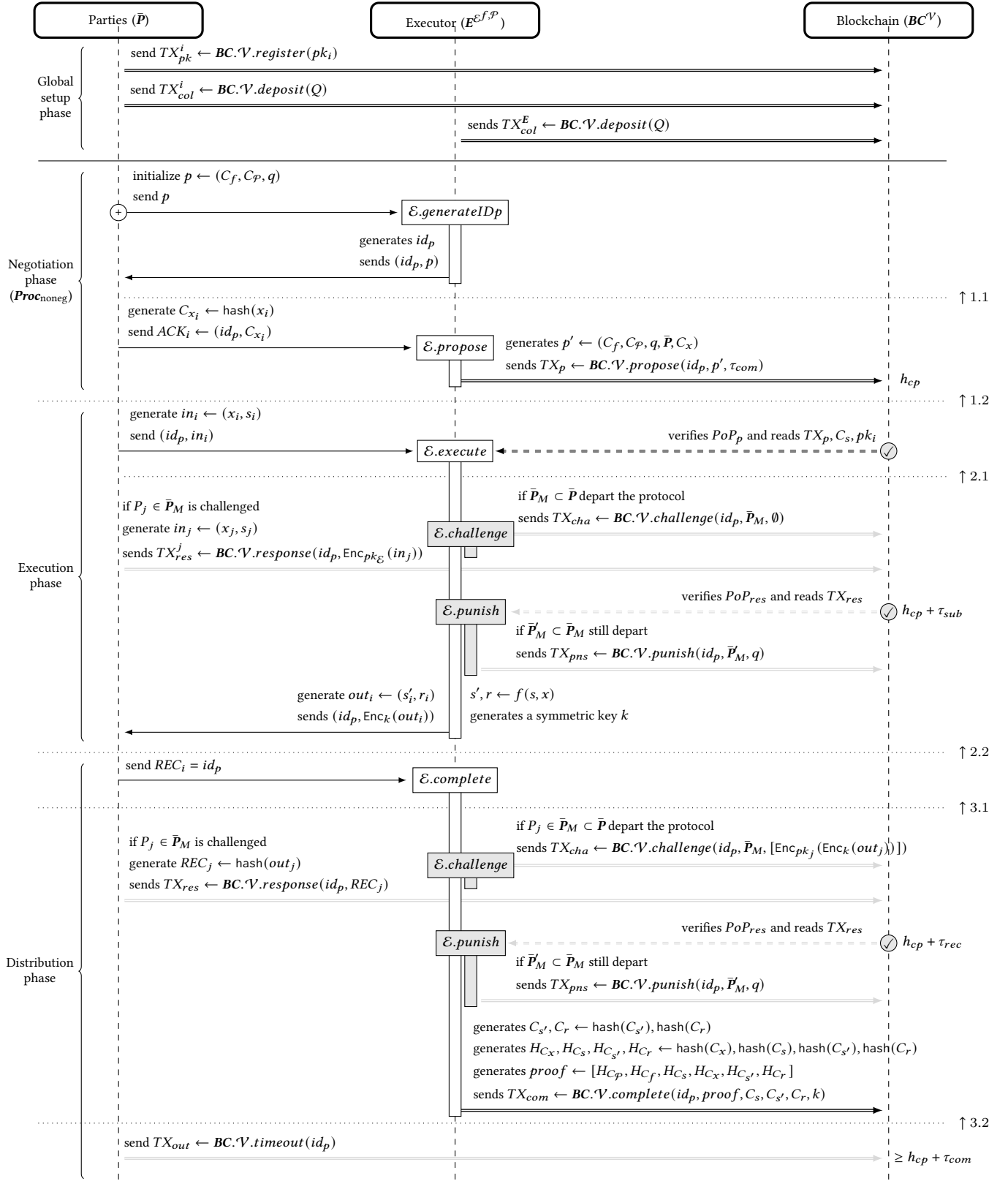
\begin{figure*}[!tbp]
    \centering
    \usetikzlibrary{calc,trees,positioning,arrows,chains,shapes.geometric,%
    decorations.pathreplacing,decorations.pathmorphing,shapes,%
    matrix,shapes.symbols, arrows.meta}
  
    \def\Initiator{Initiator}
    \def\Parties{Parties}
    \def\executor{Executor}
    \def\Blockchain{Blockchain}
    
    \def\Dividends{Dividends}
    \def\Acks{Acks}
    \def\ConfirmTx{TXs}
    \def\GetConfirmTx{GetTXs}
    \def\Proof{Proof}
    \def\PoP{PoP}
    \def\PunishExecutor{PunishExecutor}
	
    \tikzstyle{inout}=[trapezium, trapezium left angle=60, trapezium right angle=120, draw] 
    \tikzstyle{end}=[rectangle, rounded corners, draw]   
    \tikzstyle{endn}=[rounded rectangle, draw]   
    \tikzstyle{exec}=[rectangle, draw]    
    \tikzstyle{decide}=[kite, kite vertex angles=120, draw]   

    \tikzset{
    >=stealth',
      punktchain/.style={
        rectangle, 
        rounded corners, 
        draw=black, very thick,
        text width=10em, 
        minimum height=3em, 
        text centered, 
        on chain},
      line/.style={draw, thick, <-},
      element/.style={
        tape,
        top color=white,
        bottom color=blue!50!black!60!,
        minimum width=8em,
        draw=blue!40!black!90, very thick,
        text width=10em, 
        minimum height=3.5em, 
        text centered, 
        on chain},
      every join/.style={->, thick,shorten >=1pt},
      decoration={brace},
      tuborg/.style={decorate},
      tubnode/.style={midway, left=2pt},
        >=stealth',
        punkt/.style={
               rectangle,
               rounded corners,
               draw=black, very thick,
               text width=9em,
               minimum height=2em,
               text centered},
        pil/.style={
           ->,
           thick,
           shorten <=2pt,
           shorten >=2pt,}
    }
	
    \begin{tikzpicture}[every node/.style={font=\footnotesize,
      minimum height=0.1cm,minimum width=0.1cm}]
        \node [matrix, very thin, column sep=1.9cm, row sep=0.7cm] (matrix) at (0,0) {
            &[-4ex] \node(0,0) (\Parties) {}; &[14ex] & \node(0,0) (\executor) {}; &[27ex] & \node(0,0) (\Blockchain) {}; &[-7ex]  \\
            [0ex]& \node(0,0) (\Parties G0) {}; & \node(0,0) (\ConfirmTx G0p-b) {};  & \node(0,0) (\executor G0) {}; & & \node(0,0) (\Blockchain G0) {};  & \\
            [-2ex]& \node(0,0) (\Parties G1) {}; & \node(0,0) (\ConfirmTx G1p-b) {};  & \node(0,0) (\executor G1) {}; & & \node(0,0) (\Blockchain G1) {};  & \\
            [-2ex]& \node(0,0) (\Parties G2) {}; & & \node(0,0) (\executor G2) {}; & \node(0,0) (\ConfirmTx G2p-b) {};   & \node(0,0) (\Blockchain G2) {};  & \\
            [-4ex] \node(0,0) (tg1 left) {}; & \node(0,0) (\Parties tg1-1) {};  & & \node(0,0) (\executor tg1-1) {};  & & & \node(0,0) (tg1 right) {}; \\
            [-2ex] & \node(0,0) (\Parties 0) {}; &  & \node(0,0) (\executor 0) {}; & & \node(0,0) (\Blockchain 0) {};  & \\
            [-4.5ex] \node(0,0) (t0 left) {}; & \node(0,0) (\Parties 0-1) {};  & \node(0,0) (\Dividends 0-1) {}; & \node(0,0) (\executor 0-1) {};  & & & \node(0,0) (t0 right) {}; \\
            [2ex]& \node(0,0) (\Parties 1) {}; & \node(0,0) (\Dividends 1) {}; & \node(0,0) (\executor 1) {}; & & & \\
            [-4ex] \node(0,0) (t1 left) {}; &\node(0,0) (\Parties 1-1) {};  & \node(0,0) (\Dividends 1-1) {};  & \node(0,0) (\executor 1-1) {};  & & & \node(0,0) (t1 right) {};\\
            [0ex] & \node(0,0) (\Parties 2) {}; & \node(0,0) (\Acks) {}; & \node(0,0) (\executor 2) {}; & \node(0,0) (\GetConfirmTx 2) {};  & \node(0,0) (\Blockchain 2) {};  & \\
            [-10ex] \node(0,0) (t2 left) {}; & \node(0,0) (\Parties 2-1) {};  & & \node(0,0) (\executor 2-1) {};  & & \node(0,0) (\Blockchain 2-1) {};  & \node(0,0) (t2 right) {};\\ 
            [1ex] & \node(0,0) (\Parties 3) {}; & & \node(0,0) (\executor 3) {}; & \node(0,0) (\ConfirmTx proposal) {};  & \node(0,0) (\Blockchain 3) {}; & \\
            [-4ex] \node(0,0) (t3 left) {}; & \node(0,0) (\Parties 3-1) {};  & & & & & \node(0,0) (t3 right) {}; \\
            [0ex] & \node(0,0) (\Parties 4) {}; & \node(0,0) (\Dividends 4) {};  & \node(0,0) (\executor 4) {}; & \node(0,0) (\GetConfirmTx 4)  {}; & \node(0,0) (\Blockchain 4) {}; &  \\
            [-6ex] & \node(0,0) (\Parties 4-0) {}; & \node(0,0) (\Dividends 4-0) {}; & \node(0,0) (\executor 4-0) {}; &  & \node(0,0) (\Blockchain 4-0) {}; & \\
            [-4ex] \node(0,0) (t4-0-1 left) {}; & \node(0,0) (\Parties 4-0-1) {};  & & & & & \node(0,0) (t4-0-1 right) {};\\
            [1ex] \node(0,0) (t4-1 left) {}; & \node(0,0) (\Parties 4-1) {}; & & \node(0,0) (\executor 4-1) {}; & \node(0,0) (\ConfirmTx inputchallenge) {};  & \node(0,0) (\Blockchain 4-1) {}; & \node(0,0) (t4-1 right) {};  \\
            [-2ex] \node(0,0) (t4-2 left) {}; & \node(0,0) (\Parties 4-2) {}; & \node(0,0) (\ConfirmTx inputresponse) {};  & \node(0,0) (\executor 4-2) {}; &  & \node(0,0) (\Blockchain 4-2) {};  & \node(0,0) (t4-2 right) {};  \\
            [-1ex] \node(0,0) (t4-3 left) {}; & \node(0,0) (\Parties 4-3) {}; & & \node(0,0) (\executor 4-3) {}; & \node(0,0) (\GetConfirmTx inputresponse) {};  & \node(0,0) (\Blockchain 4-3) {};  & \node(0,0) (t4-3 right) {};  \\
            [1ex] & \node(0,0) (\Parties 5) {}; & & \node(0,0) (\executor 5) {}; & \node(0,0) (\ConfirmTx inputpunish) {};  & \node(0,0) (\Blockchain 5) {}; &  \\
            [-4ex] \node(0,0) (t5-1 left) {}; & \node(0,0) (\Parties 5-1) {}; & & \node(0,0) (\executor 5-1) {};  &  & \node(0,0) (\Blockchain 5-1) {}; & \node(0,0) (t5-1 right) {}; \\
            [-1ex]\node(0,0) (t5-2 left) {}; & \node(0,0) (\Parties 5-2) {}; & \node(0,0) (\Dividends 5-2) {}; & \node(0,0) (\executor 5-2) {};  &  & \node(0,0) (\Blockchain 5-2) {}; & \node(0,0) (t5-2 right) {}; \\ 
            [-4ex]\node(0,0) (t5-3 left) {}; & \node(0,0) (\Parties 5-3) {}; & & \node(0,0) (\executor 5-3) {};  &  & \node(0,0) (\Blockchain 5-3) {}; & \node(0,0) (t5-3 right) {}; \\ 
            [-3ex] & \node(0,0) (\Parties 7) {}; & \node(0,0) (\Dividends 7) {}; & \node(0,0) (\executor 7) {}; &  & \node(0,0) (\Blockchain 7) {}; & \\
            [-3ex] \node(0,0) (t7- left) {}; & & & & & & \node(0,0) (t7- right) {};\\
            [1ex] & \node(0,0) (\Parties 7-0) {}; & & \node(0,0) (\executor 7-0) {}; & \node(0,0) (\ConfirmTx outputchallenge) {};  & \node(0,0) (\Blockchain 7-0) {}; & \node(0,0) (t7-0 right) {};  \\
            [-2ex] & \node(0,0) (\Parties 7-1) {}; & \node(0,0) (\ConfirmTx outputresponse) {};  & \node(0,0) (\executor 7-1) {}; &  & \node(0,0) (\Blockchain 7-1) {};  & \node(0,0) (t7-1 right) {};  \\
            [-1ex] & \node(0,0) (\Parties 7-2) {}; & & \node(0,0) (\executor 7-2) {}; & \node(0,0) (\GetConfirmTx outputresponse) {};  & \node(0,0) (\Blockchain 7-2) {};  & \node(0,0) (t7-2 right) {};  \\
            [1ex] & \node(0,0) (\Parties 7-3) {}; & & \node(0,0) (\executor 7-3) {}; & \node(0,0) (\ConfirmTx outputpunish) {};  & \node(0,0) (\Blockchain 7-3) {}; &  \\
            [8ex] & \node(0,0) (\Parties 7-4) {};  & \node(0,0) () {};  & \node(0,0) (\executor 7-4) {}; & \node(0,0) (\Proof 7-4) {};  & \node(0,0) (\Blockchain 7-4) {};  & \\
            [-5ex] \node(0,0) (t7 left) {}; & & & & & & \node(0,0) (t7 right) {};\\
            [-2ex]  & \node(0,0) (\Parties 7-5) {};  & \node(0,0) (\PunishExecutor) {};  & \node(0,0) (\executor 7-5) {}; &  & \node(0,0) (\Blockchain 7-5) {};  & \\
            [-4ex] & \node(0,0) (\Parties 9) {}; & & \node(0,0) (\executor 9) {}; & & \node(0,0) (\Blockchain 9) {}; & \\
        };
        
        \fill 
        	(\Parties) node[punkt] {\Parties~($\bar{\textbf{\textit{P}}}$)}
        	(\executor) node[punkt] {\executor~($\textbf{\textit{E}}^{\mathcal{E}^{f, \mathcal{P}}}$)}
            (\Blockchain) node[punkt] {\Blockchain~($\textbf{\textit{BC}}^{\mathcal{V}}$)};
            
        \draw [dotted] 
          (\Parties 1-1) -- (t1 right) node[right] {$\uparrow 1.1$}
          (t3 left) -- (t3 right) node[right] {$\uparrow 1.2$}
          (\Parties 4-0-1) -- (t4-0-1 right) node[right] {$\uparrow 2.1$}
          (t5-3 left) -- (t5-3 right) node[right,rotate=0] {$\uparrow 2.2$}
          (t7- left) -- (t7- right) node[right,rotate=0] {$\uparrow 3.1$}
          (t7 left) -- (t7 right) node[right,rotate=0] {$\uparrow 3.2$};
          
        \draw [dashed] 
          (\Parties) -- (\Parties 9.south)
          (\executor) -- (\executor 9.south)
          (\Blockchain) -- (\Blockchain 9.south);
        
        \draw
            (tg1 left) -- (tg1 right) node[right] {};

        \draw [-latex] let \p1=(\Parties 0-1), \p2=(\executor 0-1) in ($(\x1, \y1)$) -- ($(\x2-3.3em, \y2)$);
        \draw [-latex] (\executor 1.west) -- (\Parties 1);
        \draw [-latex] let \p1=(\Parties 2), \p2=(\executor 2) in ($(\x1, \y1)$) -- ($(\x2-2.5em, \y2)$);
        \draw [-latex] let \p1=(\Parties 4), \p2=(\executor 4) in ($(\x1, \y1)$) -- ($(\x2-2.4em, \y2)$);
        \draw [-latex] (\executor 5-2) -- (\Parties 5-2);
        \draw [-latex] let \p1=(\Parties 7), \p2=(\executor 7) in ($(\x1, \y1)$) -- ($(\x2-2.8em, \y2)$);
        
        
        \draw [line width=0.5pt, double distance=0.5pt, arrows = {-Latex[length=1.5pt 2 0]}] (\Parties G0) -- (\Blockchain G0);
        \draw [line width=0.5pt, double distance=0.5pt, arrows = {-Latex[length=1.5pt 2 0]}] (\Parties G1) -- (\Blockchain G1);
        \draw [line width=0.5pt, double distance=0.5pt, arrows = {-Latex[length=1.5pt 2 0]}] (\executor G2) -- (\Blockchain G2);
        \draw [line width=0.5pt, double distance=0.5pt, arrows = {-Latex[length=1.5pt 2 0]}] (\executor 3) -- (\Blockchain 3);
        \draw [line width=0.5pt, double distance=0.5pt, arrows = {-latex}, color=gray!30] let \p1=(\executor 4-1), \p2=(\Blockchain 4-1.west) in ($(\x1+2.5em, \y1)$) -- ($(\x2, \y2)$); 
        \draw [line width=0.5pt, double distance=0.5pt, arrows = {-latex}, color=gray!30] (\Parties 4-2) -- (\Blockchain 4-2);
        \draw [line width=0.5pt, double distance=0.5pt, arrows = {-latex}, color=gray!30] let \p1=(\executor 5), \p2=(\Blockchain 5.west) in ($(\x1+1.5em, \y1)$) -- ($(\x2, \y2)$); 
        \draw [line width=0.5pt, double distance=0.5pt, arrows = {-latex}, color=gray!30] let \p1=(\executor 7-0), \p2=(\Blockchain 7-0.west) in ($(\x1+2.5em, \y1)$) -- ($(\x2, \y2)$); 
        \draw [line width=0.5pt, double distance=0.5pt, arrows = {-latex}, color=gray!30] (\Parties 7-1) -- (\Blockchain 7-1);
        \draw [line width=0.5pt, double distance=0.5pt, arrows = {-latex}, color=gray!30] let \p1=(\executor 7-3), \p2=(\Blockchain 7-3.west) in ($(\x1+1.5em, \y1)$) -- ($(\x2, \y2)$); 
        \draw [line width=0.5pt, double distance=0.5pt, arrows = {-latex}] (\executor 7-4) -- (\Blockchain 7-4);
        \draw [line width=0.5pt, double distance=0.5pt, arrows = {-latex}, color=gray!30] (\Parties 7-5) -- (\Blockchain 7-5);
        
        \draw [dashed, double distance=0.5pt, arrows = {-latex}] let \p1=(\Blockchain 4), \p2=(\executor 4) in ($(\x1, \y1)$) -- ($(\x2+2.3em, \y2)$);
        \draw [dashed, double distance=0.5pt, arrows = {-latex}, color=gray!30] let \p1=(\Blockchain 4-3), \p2=(\executor 4-3) in ($(\x1, \y1)$) -- ($(\x2+3.2em, \y2)$); 
        \draw [dashed, double distance=0.5pt, arrows = {-latex}, color=gray!30] let \p1=(\Blockchain 7-2), \p2=(\executor 7-2) in ($(\x1, \y1)$) -- ($(\x2+3.0em, \y2)$); 
        
        \draw[tuborg, decoration={brace}] let \p1=(\Parties tg1-1.south), \p2=(\Parties G0.north) in ($(\x1-2em, \y1+1em)$) -- ($(\x1-2em, \y2+1em)$) node[tubnode] {\makecell[c]{Global\\setup\\phase}};
        \draw[tuborg, decoration={brace}] let \p1=(\Parties 3-1.south), \p2=(\Parties tg1-1.north) in ($(\x1-2em, \y1+1em)$) -- ($(\x1-2em, \y2-1em)$) node[tubnode] {\makecell[c]{Negotiation\\phase\\($\textbf{\textit{Proc}}_\text{noneg}$)}};
        \draw[tuborg, decoration={brace}] let \p1=(\Parties 5-3.south), \p2=(\Parties 3-1.north) in ($(\x1-2em, \y1+1em)$) -- ($(\x1-2em, \y2-1em)$) node[tubnode] {\makecell[c]{Execution\\phase}};
        \draw[tuborg, decoration={brace}] let \p1=(\Parties 9), \p2=(\Parties 5-3.north) in ($(\x1-2em, \y1)$) -- ($(\x1-2em, \y2-1em)$) node[tubnode] {\makecell[c]{Distribution\\phase\\}};

        \fill
            (\ConfirmTx G0p-b) 
                node[above, shift={(0:-1.9cm)}] {
                    \begin{math}
                        \begin{aligned}
                            \text{send}~TX^i_{pk}\gets\textbf{\textit{BC}}.\mathcal{V}.register(pk_i)
                        \end{aligned}
                    \end{math}
                }
                node[font=\footnotesize, below] {}
            (\ConfirmTx G1p-b) 
                node[above, shift={(0:-2.0cm)}] {
                    \begin{math}
                        \begin{aligned}
                            \text{send}~TX^i_{col}\gets\textbf{\textit{BC}}.\mathcal{V}.deposit(Q)
                        \end{aligned}
                    \end{math}
                }
                node[font=\footnotesize, below] {}
            (\ConfirmTx G2p-b) 
                node[above, shift={(0:-3.7cm)}] {
                    \begin{math}
                        \begin{aligned}
                            \text{sends}~TX^{\textit{\textbf{E}}}_{col}\gets\textbf{\textit{BC}}.\mathcal{V}.deposit(Q)
                        \end{aligned}
                    \end{math}
                }
                node[font=\footnotesize, below] {}
            (\Dividends 0-1)  
                node[above, shift={(0:-2.4cm)}] {
                    \begin{math}
                        \begin{aligned}
                            & \text{initialize}~p\gets(C_f, C_{\mathcal{P}},  q) \\
                            & \text{send}~p
                        \end{aligned}
                    \end{math}
                }
                node[font=\footnotesize, below] {}
            (\Dividends 1) 
                node[above, shift={(0:1.1cm)}] {
                    \begin{math}
                        \begin{aligned}
                            & \text{generates}~id_p \\
                            & \text{sends}~( id_p, p )
                        \end{aligned}
                    \end{math}
                }
                node[font=\footnotesize, below] {}
            (\Acks) 
                node[above, shift={(0:-2.5cm)}] {
                    \begin{math}
                        \begin{aligned}
                            & \text{generate}~C_{x_i}\gets \FuncSty{hash}(x_i) \\
                            & \text{send}~ACK_i\gets( id_p, C_{x_i})
                        \end{aligned}
                    \end{math}
                }
                node[font=\footnotesize, below] {}
            (\ConfirmTx proposal) 
                node[above, shift={(0:-2.5cm)}] {
                    \begin{math}
                        \begin{aligned}
                            & \text{generates}~p'\gets(C_f, C_{\mathcal{P}},  q, \bar{\textit{\textbf{P}}}, C_{x})\\
                            & \text{sends}~TX_p\gets\textbf{\textit{BC}}.\mathcal{V}.propose(id_p, p', \tau_{com})
                        \end{aligned}
                    \end{math}
                }
                node[font=\footnotesize, below] {}
            (\Blockchain 3) 
                node[shift={(0:0.4cm)}] {
                    $h_{cp}$
                }
            (\ConfirmTx inputchallenge) 
                node[above, shift={(0:-2.2cm)}] {
                    \begin{math}
                        \begin{aligned}
                            & \text{if}~\bar{\textit{\textbf{P}}}_M\subset \bar{\textit{\textbf{P}}} ~\text{depart the protocol} \\
                            & \text{sends}~TX_{cha}\gets\textbf{\textit{BC}}.\mathcal{V}.challenge(id_p, \bar{\textit{\textbf{P}}}_M, \emptyset ) 
                        \end{aligned}
                    \end{math}
                }
                node[font=\footnotesize, below] {}
            (\ConfirmTx inputresponse) 
                node[above, shift={(0:-1.2cm)}] {
                    \begin{math}
                        \begin{aligned}
                            & \text{if}~P_j\in \bar{\textit{\textbf{P}}}_M~\text{is challenged} \\
                            & \text{generate}~in_j\gets( x_j, s_j ) \\
                            & \text{sends}~TX^j_{res}\gets\textbf{\textit{BC}}.\mathcal{V}.response(id_p, \FuncSty{Enc}_{pk_{\mathcal{E}}}(in_j))
                        \end{aligned}
                    \end{math}
                }
                node[font=\footnotesize, below] {}
            (\GetConfirmTx inputresponse) 
                node[above, shift={(0:0cm)}] {
                    \begin{math}
                        \begin{aligned}
                            & \text{verifies}~PoP_{res}~\text{and reads}~TX_{res}\\
                        \end{aligned}
                    \end{math}
                }
                node[font=\footnotesize, below] {}    
            (\ConfirmTx inputpunish) 
                node[above, shift={(0:-2.4cm)}] {
                    \begin{math}
                        \begin{aligned}
                            & \text{if}~\bar{\textit{\textbf{P}}}'_M\subset\bar{\textit{\textbf{P}}}_M~\text{still depart} \\
                            & \text{sends}~TX_{pns}\gets\textbf{\textit{BC}}.\mathcal{V}.punish(id_p, \bar{\textit{\textbf{P}}}'_M, q)\\
                        \end{aligned}
                    \end{math}
                }
                node[font=\footnotesize, below] {}
            (\Dividends 4)  
                node[above, shift={(0:-2.6cm)}] {
                    \begin{math}
                        \begin{aligned}
                            & \text{generate}~in_i\gets( x_i, s_i)\\
                            & \text{send}~( id_p, in_i )
                        \end{aligned}
                    \end{math}
                }
                node[font=\footnotesize, below] {}
            (\GetConfirmTx 4)  
                node[above] {
                    \begin{math}
                        \begin{aligned}
                            \text{verifies}~PoP_{p}~\text{and reads}~TX_p, C_s, pk_i
                        \end{aligned}
                    \end{math}
                }
                node[font=\footnotesize, below] {}
            (\Blockchain 4-3) 
                node[shift={(0:0.8cm)}] {
                    $h_{cp}+\tau_{sub}$
                }
            (\executor 5-2)
                node[right, shift={(2.5cm:0.5cm)}] {
                    \begin{math}
                        \begin{aligned}
                            & s', r \gets f(s, x) \\
                            & \text{generates a symmetric key}~k \\
                        \end{aligned}
                    \end{math}
                }
            (\Dividends 5-2) 
                node[above, shift={(0:0.5cm)}] {
                    \begin{math}
                        \begin{aligned}
                            & \text{generate}~out_i \gets ( s'_i, r_i) \\
                            & \text{sends}~ ( id_p, \FuncSty{Enc}_{k}(out_i)) 
                        \end{aligned}
                    \end{math}
                }
                node[font=\footnotesize, below] {}
            (\Dividends 7) 
                node[above, shift={(0:-3.0cm)}] {
                    \begin{math}
                        \text{send}~REC_i=id_p
                    \end{math}
                }
                node[font=\footnotesize, below] {}
            (\ConfirmTx outputchallenge) 
                node[above, shift={(0:-1.1cm)}] {
                    \begin{math}
                        \begin{aligned}
                            & \text{if}~P_j\in \bar{\textit{\textbf{P}}}_M\subset \bar{\textit{\textbf{P}}} ~\text{depart the protocol} \\
                            & \text{sends}~TX_{cha}\gets\textbf{\textit{BC}}.\mathcal{V}.challenge(id_p, \bar{\textit{\textbf{P}}}_M, [\FuncSty{Enc}_{pk_{j}}(\FuncSty{Enc}_{k}(out_j))])
                        \end{aligned}
                    \end{math}
                }
                node[font=\footnotesize, below] {}
            (\ConfirmTx outputresponse) 
                node[above, shift={(0:-1.45cm)}] {
                    \begin{math}
                        \begin{aligned}
                            & \text{if}~P_j\in \bar{\textit{\textbf{P}}}_M~\text{is challenged} \\
                            & \text{generate}~REC_j\gets \FuncSty{hash}(out_j) \\
                            & \text{sends}~TX_{res}\gets\textbf{\textit{BC}}.\mathcal{V}.response(id_p, REC_j)
                        \end{aligned}
                    \end{math}
                }
                node[font=\footnotesize, below] {}
            (\GetConfirmTx outputresponse) 
                node[above, shift={(0:0cm)}] {
                    \begin{math}
                        \begin{aligned}
                            & \text{verifies}~PoP_{res}~\text{and reads}~TX_{res}\\
                        \end{aligned}
                    \end{math}
                }
                node[font=\footnotesize, below] {}    
            (\ConfirmTx outputpunish) 
                node[above, shift={(0:-2.4cm)}] {
                    \begin{math}
                        \begin{aligned}
                            & \text{if}~\bar{\textit{\textbf{P}}}'_M\subset\bar{\textit{\textbf{P}}}_M~\text{still depart} \\
                            & \text{sends}~TX_{pns}\gets\textbf{\textit{BC}}.\mathcal{V}.punish(id_p, \bar{\textit{\textbf{P}}}'_M, q)\\
                        \end{aligned}
                    \end{math}
                }
                node[font=\footnotesize, below] {}
            (\Blockchain 7-2) 
                node[shift={(0:0.8cm)}] {
                    $h_{cp}+\tau_{rec}$
                }
            (\Proof 7-4) 
                node[above, shift={(0cm:-1.4cm)}] {
                    \begin{math}
                        \begin{aligned}
                            & \text{generates}~C_{s'}, C_{r} \gets \FuncSty{hash}(C_{s'}), \FuncSty{hash}(C_{r}) \\
                            & \text{generates}~H_{C_x}, H_{C_s}, H_{C_{s'}}, H_{C_{r}} \gets \FuncSty{hash}(C_x), \FuncSty{hash}(C_s), \FuncSty{hash}(C_{s'}), \FuncSty{hash}(C_r) \\
                            & \text{generates}~proof\gets[H_{C_{\mathcal{P}}}, H_{C_f}, H_{C_s}, H_{C_x}, H_{C_{s'}}, H_{C_r}] \\
                            &\text{sends}~TX_{com}\gets\textbf{\textit{BC}}.\mathcal{V}.complete(id_p, proof, C_{s}, C_{s'}, C_r, k)
                        \end{aligned}
                    \end{math}
                }
                node[below] {}
            (\PunishExecutor) 
                node[above, shift={(0:-1.8cm)}] {
                    \begin{math}
                        \begin{aligned}
                          & \text{send}~TX_{out}\gets\textbf{\textit{BC}}.\mathcal{V}.timeout(id_p)
                        \end{aligned}
                    \end{math}
                }
                node[font=\footnotesize, below] {}
            (\Blockchain 7-5) 
                node[shift={(0:0.8cm)}] {
                    $\geq h_{cp} + \tau_{com}$
                }
                node[font=\footnotesize, below] {};
        \filldraw[fill=white]
            (\executor 0-1.north west) rectangle (\executor 1.south east)
            (\executor 2.north west) rectangle (\executor 3.south east)
            (\executor 4.north west) rectangle (\executor 5-2.south east)
            (\executor 7.north west) rectangle (\executor 7-4.south east);
    
        \filldraw[fill=gray!20]
            let \p1=(\executor 4-1.north west), \p2=(\executor 4-1.south east) in ($(\x1+0.9em, \y1)$) rectangle ($(\x2+0.9em, \y2-1em)$)
            let \p1=(\executor 4-3.north west), \p2=(\executor 5.south east) in ($(\x1+0.9em, \y1)$) rectangle ($(\x2+0.9em, \y2)$)
            let \p1=(\executor 7-0.north west), \p2=(\executor 7-0.south east) in ($(\x1+0.9em, \y1)$) rectangle ($(\x2+0.9em, \y2-1em)$)
            let \p1=(\executor 7-2.north west), \p2=(\executor 7-3.south east) in ($(\x1+0.9em, \y1)$) rectangle ($(\x2+0.9em, \y2)$);
            
        \draw
            (\Parties 0-1) node[draw,circle,fill=white!20] {} node {+}
            (\executor 0-1) node[draw, fill=white] {$\mathcal{E}.generateIDp$}
            (\executor 2) node[draw, fill=white] {$\mathcal{E}.propose$}
            (\executor 4) node[draw, fill=white] {$\mathcal{E}.execute$}
            (\executor 4-1) node[draw, fill=gray!20, xshift=0.9em] {$\mathcal{E}.challenge$}
            (\executor 4-3) node[draw, fill=gray!20, xshift=0.9em] {$\mathcal{E}.punish$}
            (\executor 7) node[draw, fill=white] {$\mathcal{E}.complete$}
            (\executor 7-0) node[draw, fill=gray!20, xshift=0.9em] {$\mathcal{E}.challenge$}
            (\executor 7-2) node[draw, fill=gray!20, xshift=0.9em] {$\mathcal{E}.punish$}
            
            (\Blockchain 4) node[draw,circle,fill=gray!20] {} node {\checkmark}
            (\Blockchain 4-3) node[draw,circle,fill=gray!20] {} node {\checkmark}
            (\Blockchain 7-2) node[draw,circle,fill=gray!20] {} node {\checkmark};

    \end{tikzpicture}
    
    \caption{\textbf{Detailed \codename protocol $\pi_{\codename}$}. \small{$\bar{\bm P}$ refers to parties participating in an \acrshort{mpt}. ${\bm E}^{\mathcal{E}^{f, \mathcal{P}}}$ refers to an executor holding a \acrshort{tee} enclave $\mathcal{E}$ with deployed $f, \mathcal{P}$. ${\bm BC}^{\mathcal{V}}$ refers to the blockchain with deployed contract program $\mathcal{V}$, where $\mathcal{V}$ is an on-chain verifier that manages the life cycle of \acrshort{mpt} and accepts output commitments by verifying the state transition proved by $proof$. Double dashed arrows refer to reading from the blockchain and double arrows refer to sending a transaction to the blockchain. Normal arrows indicate off-chain communication. \textbf{\textit{Proc}}$_\text{noneg}$ refers to nondeterministic negotiation protocol. All parties $P_i$ communicate with the executor in secure channels; thus, we omit marking ciphertext in communications between $\bar{\bm P}$ and $\mathcal{E}$ for simplicity but explicitly mark the ciphertext in transactions published on ${\bm BC}^{\mathcal{V}}$. } } 
    \label{prot:cloak}
\end{figure*}

In this section, we illustrate how \codename protocol $\pi_{\codename}$ enforces \acrshort{mpt} in detail. Given a blockchain \textbf{\textit{BC}}, a party set $\bar{\bm P} (|\bar{\bm P}|=n$) participating the \acrshort{mpt}, and an executor \textbf{\textit{E}}, Figure~\ref{prot:cloak} depicts the detailed phases and messages of the \codename protocol. Each $P_i\in\bar{\bm P}$ communicates with $\mathcal{E}$ by secure channels\footnote{Messages sent from a party $P_i$ to \textbf{\textit{E}} are signed by $P_i$ and encrypted by $pk_{\mathcal{E}}$, while messages sent from $\mathcal{E}$ to $P_i$ are also signed by $\mathcal{E}$.}. For simplicity, we only mark the ciphertext not for building secure channels, \eg, the ciphertext in each transaction sent to the blockchain.

As described in Section~\ref{sec:overview}, our protocol $\pi_{\codename}$ proceeds in four phases. To summarise, before sending an \acrshort{mpt}, \textit{\textbf{E}} and each party are required to $deposit$ some coins $Q$ in the \textit{global setup phase}. Both subjects go through the setup phase only once. Then, three phases follow to evaluate an \acrshort{mpt}. During a \textit{negotiation phase}, all parties negotiate off-chain to join the \acrshort{mpt}, and finally, ${\bm E}$ commits a settled \acrshort{mpt} proposal with parties' input commitments and deducts their collateral on-chain. Next, an \textit{execution phase} follows for collecting plaintext of parameters and old states from parties and executing \acrshort{mpt} in the enclave to obtain outputs and deliver the output ciphertext to parties. Finally, the protocol enters a \textit{distribution phase} to commit outputs and reveal the encryption key to complete the \acrshort{mpt}. We now explain the detailed protocol phases. Protocol security parameters such as $t_*, \tau_*$ are quantified in Appendix~\ref{sec:security-proof}.

\vspace{-0.15cm}
\subsection{Negotiation phase} 
This phase uses a \textit{nondeterministic negotiation protocol} (\textbf{\textbf{\textit{Proc}}}$_\text{noneg}$) to guide parties to reach a consensus on an \acrshort{mpt} proposal and commit parameters $x_i$ on-chain\footnote{The old state $s_i$ is already committed on-chain before starting this \acrshort{mpt}}. \textbf{{\textit{Proc}}}$_\text{noneg}$ proceeds in two stages. 

\textbf{\textit{1.1}}: One party wishing to call an \acrshort{mpt} sends an \acrshort{mpt} proposal $p \gets ( C_f, C_{\mathcal{P}}, q)$ to $\mathcal{E}$, where $q$ refers to required collateral for punishing adversary. Then, $\mathcal{E}$ generates a random identifier for proposal $p$, $id_p$, and broadcasts the signed $( id_p, p)$ to parties $\bar{\bm P}$. After receiving $( id_p, p)$, each $P_i$ interested in the \acrshort{mpt} computes the commitment $C_{x_i}$ of its parameter $x_i$ and sends a signed acknowledgment $ACK_i\gets( id_p, C_{x_i})$ to $\mathcal{E}$ before $t_n$, where $t_n$ refers to the negotiation phase's completion time. $\mathcal{E}$ knows that $P_i$ is interested in \acrshort{mpt} when it receives $ACK_i$ . 

\textbf{\textit{1.2}}: If the collected acknowledgements satisfy the settlement condition\footnote{Settlement conditions of negotiation can be specified in $\mathcal{P}$, \eg, requiring specific parties joining \acrshort{mpt} or the number of parties exceeding a specific number.} specified in $p$, $\mathcal{E}$ creates a settled proposal $p'$. $p'$ expands $p$ by adding the addresses of parties $\bar{\bm P}$ and the array containing all parties' parameter commitments $C_x$. Then, $\mathcal{E}$ sends $TX_p$ to the blockchain to confirm $p'$. Additionally, $TX_p$ deducts $q$ collateral from each party and $n*q$ from the executor prior to executing \acrshort{mpt}. Finally, $\mathcal{E}$ enters the \textit{execution phase}. Otherwise, if the settlement condition is not satisfied and the duration exceeds $t_n$, the negotiation of $p$ fails, and $\mathcal{E}$ terminates the protocol.

\vspace{-0.15cm}
\subsection{Execution phase}
This phase collects plaintext inputs from parties and evaluates the \acrshort{mpt} to obtain outputs. It normally contains two stages. When some subjects misbehave, an additionally \emph{challenge-response submission} stage arises.

\textbf{\textit{2.1}}: Upon confirmation of $TX_p$ on \textit{\textbf{BC}}, $\mathcal{E}$ reads the view of \textit{\textbf{BC}} in order to validate the \acrshort{pop}~\cite{Ekiden:2019, Cavallaro2019Tesseract} of $TX_p$, \ie, ($PoP_{p}$), and reads the old state commitments $C_s$ from \textit{\textbf{BC}}. Additionally, any party $P_i$ that is aware of its involvement in $TX_p$ submits inputs (\ie, parameters $x_i$ and old states $s_i$) to $\mathcal{E}$. $\mathcal{E}$, upon receiving inputs from $P_i$, recomputes commitments of $x_i, s_i$ in order to match them to their corresponding commitments $C_{x_i}, C_{s_i}$ on \textit{\textbf{BC}}. If all inputs from all parties are collected and matched, $\mathcal{E}$ proceeds to \textbf{\textit{2.2}}. Otherwise, if $\mathcal{E}$ discovers that certain parties' inputs conflict with their on-chain commitments or that some parties' inputs are not received before $t_e$, $\mathcal{E}$ flags these parties as potentially misbehaving and returns $\bar{\bm P}_M$ to \textbf{\textit{E}}. Then, \textbf{\textit{E}} invokes $\mathcal{E}.challenge$ to send $TX_{cha}$. $TX_{cha}$ publicly challenges all parties in $\bar{\bm P}_M$ on-chain. Following that, $\mathcal{E}$ proceeds to the \textbf{\textit{challenge-response submission}} stage.

\textbf{\textit{challenge-response submission}}: After $TX_{cha}$ is confirmed on-chain, parties belong to $\bar{\bm P}_M$ but are honest send a $TX_{res}$ to publish ciphertext of their inputs $x_i, s_i$. All published $TX_{res}$ must be confirmed prior to the block $h_{cp}+\tau_{sub}$. After all published $TX_{res}$ are confirmed, $\mathcal{E}$ reads published $TX_{res}$ and verifies $PoP_{res}$ (recall that \acrshort{pop} is for proving a message has been confirmed on the blockchain). If $\mathcal{E}$ successfully reads matching inputs from $TX^i_{res}$, $\mathcal{E}$ deletes $P_i$ from $\bar{\bm P}_M$. Otherwise, if $PoP_{res}$ shows that no $TX^i_{res}$ of $P_i \in \bar{\bm P}_M$ has been published or the inputs in $TX^i_{res}$ are still mismatched, $\mathcal{E}$ maintains $P_i$ in $\bar{\bm P}_M$. Subsequently, if $\bar{\bm P}_M$ becomes empty after the challenge-response stage, indicating that all inputs have been collected, $\mathcal{E}$ proceeds to \textbf{\textit{2.2}}. Conversely, if $\bar{\bm P}_M$ remains nonempty, indicating that the misbehavior of remaining parties have been proven, $\mathcal{E}$ flags the remain parties as misbehaving parties $\bar{\bm P}'_M$. After that, $\mathcal{E}$ sends $TX_{pns}$. $TX_{pns}$ refunds all parties' deducted collateral only to honest parties on average and terminates the \acrshort{mpt} with \code{ABORT}.

\textbf{\textit{2.2}}: If all parties' inputs are correctly collected, $\mathcal{E}$ replaces the state of the EVM inside the enclave with old state $s$, then runs $f(x)$ based on $s$ to obtain \acrshort{mpt} outputs, \ie, return values $r$ and new state $s'$. Following that, $\mathcal{E}$ generates a one-time symmetric key $k$ and use it to encrypts $r$ and $s'$ to obtain their ciphertext. Finally, $\mathcal{E}$ delivers the ciphertext $Enc_{k}(s'_i, r_i)$ to each $P_i$ correspondingly.

\vspace{-0.15cm}
\subsection{Distribution phase}
Briefly, this phase aims to publish the encryption key $k$ and commit the outputs to the blockchain after ensuring that all parties have received the output ciphertext encrypted by $k$. It normally contains two stages. When some subjects misbehave, an additionally \emph{challenge-response delivery} stage arises.

\textbf{\textit{3.1}}: $\mathcal{E}$ waits for receipts the of output ciphertext from all parties. If all receipts are collected, $\mathcal{E}$ proceeds to \textbf{\textit{3.2}}. Alternatively, similar to \emph{challenge-response submission}, if $\mathcal{E}$ discovers that certain parties' receipts are invalid or have not received some parties' receipts prior to $t_d$, $\mathcal{E}$ flags these parties as potentially misbehaving parties and returns $\bar{\bm P}_M$ to \textbf{\textit{E}}. Then, \textbf{\textit{E}} calls $\mathcal{E}.challenge$ to send $TX_{cha}$. $TX_{cha}$ challenges all parties in $\bar{\bm P}_M$ on-chain with their output ciphertext additionally encrypted by their own public keys. Subsequently, $\mathcal{E}$ proceeds to the \textbf{\textit{challenge-response delivery}} stage.

\textbf{\textit{challenge-response delivery}}: After $TX_{cha}$ is confirmed, parties present in $\bar{\bm P}_M$ but are honest send $TX_{res}$ to publish their receipts of output $s'_i, r_i$ on \textbf{\textit{BC}}. All published $TX_{res}$ must be confirmed prior to the block $h_{cp}+\tau_{rec}$. When all published $TX_{res}$ have been confirmed, $\mathcal{E}$ reads the published $TX_{res}$ and verifies its $PoP_{res}$. For $P_i$ in $\bar{\bm P}_M$, if $\mathcal{E}$ successfully reads a party's receipt $REC_i$ from its $TX^i_{res}$, $\mathcal{E}$ deletes $P_i$ from $\bar{\bm P}_M$. Otherwise, if $PoP_{res}$ shows that no $TX^i_{res}$ has been published, $\mathcal{E}$ maintains $P_i$ in $\bar{\bm P}_M$. Subsequently, if $\bar{\bm P}_M$ becomes empty after this challenge-response stage, indicating that all parties' receipts have been collected, $\mathcal{E}$ proceeds to \textbf{\textit{3.2}}. Conversely, if $\bar{\bm P}_M$ remains nonempty, proving that the remaining parties misbehave, $\mathcal{E}$ flags these misbehaving parties as $\bar{\bm P}'_M$. After that, $\mathcal{E}$ sends $TX_{pns}$. This transaction refunds all parties' deducted collateral only to honest parties on average and terminates the \acrshort{mpt} with \code{ABORT}.

\textbf{\textit{3.2}}: $\mathcal{E}$ publishes $k$ and the commitments of $s'_i, r_i$ on-chain. Specifically, $\mathcal{E}$ computes the commitments of $s'_i, r_i$, yielding $C_{s'_i}, C_{r_i}$. Then, $\mathcal{E}$ sends $TX_{com}$ to the blockchain along with $k$ and $proof$. $proof\gets[H_{C_{\mathcal{P}}}, H_{C_f}, H_{C_s}, H_{C_x}, H_{C_{s'}}, H_{C_r}]$, where $H_{C_{\mathcal{P}}}$ (resp. $H_{C_f}$) denotes the commitment hash of $\mathcal{P}$ (resp. $f$) and $H_{C_{s}}$ denotes \aptLtoX[graphic=no,type=env]{$\texttt{hash}([C_{s_i}|_{1..n}])$}{$\FuncSty{hash}([C_{s_i}|_{1..n}])$}. $proof$ in $TX_{com}$ achieves public verifiability of \acrshort{mpt} in the following way: The signed $TX_{com}$ indicates that $\mathcal{E}$ endorses that by enforcing a privacy policy $\mathcal{P}$ (matching $H_{C_{\mathcal{P}}}$), it takes private $x, s$ (matching $H_{C_{x}}, H_{C_s}$) from $\bar{\bm P}$, evaluates $f$ (matching $H_{C_f}$), and obtains private outputs $s', r$ (matching $H_{C_{s'}}, H_{C_r}$). Therefore, by trusting the integrity and confidentiality of $\mathcal{E}$, if \textbf{\textit{BC}} verifies that $H_{C_{\mathcal{P}}}, H_{C_f}$ in $proof$ matches the previously registered $H_{C_{\mathcal{P}}}, H_{C_f}$, and $H_{C_{s}}$ in $proof$ matches the existing state commitments, \textbf{\textit{BC}} then accepts the state transition, updates its existing state commitments with $C_{s'}$, and signals \code{COMPLETE} of the \acrshort{mpt}. Otherwise, if $\mathcal{E}$ neither completes (via $TX_{com}$) or terminates (via $TX_{pns}$) the \acrshort{mpt} before the block height $\tau_{com}$, it indicates that \textbf{\textit{E}} misbehaves. Then, any $P_i$ can send $TX_{out}$ to punish \textbf{\textit{E}} and refund their collateral with additional compensation. $TX_{out}$ also terminates \acrshort{mpt} with \code{ABORT}. 
    
\vspace{-0.15cm}  
\section{Implementation} \label{sec:implementation}
    For prototyping, \codename instantiates the \acrshort{tee} as Intel SGX and the blockchain as Ethereum. 

\vspace{-0.15cm}
\subsection{Contract facility}
We use Ganache~\cite{ganache} to simulate an legacy Ethereum, \ie, the \textbf{\textit{BC}} in \codename. To enable the \acrshort{mpt} on existing \textbf{\textit{BC}}, we require the executor to deploy a contract program $\mathcal{V}$ to \textbf{\textit{BC}}. As is shown in Algorithm~\ref{alg:cloak-service}, $\mathcal{V}$ is constructed by the config of $\mathcal{E}$, \eg, $pk_{\mathcal{E}}$ and $adr_{\mathcal{E}}$, so that parties can attest the integrity of $\mathcal{E}$ by its IAS report and then build secure channels with $\mathcal{E}$ by $pk_{\mathcal{E}}$. Moreover, $\mathcal{V}$ provides several functions to manage life cycles of \acrshort{mpt}.

\begin{algorithm}[!htbp]
    \footnotesize
    \DontPrintSemicolon
    \LinesNumbered
    \caption{Contract program ($\mathcal{V}$)}
    \label{alg:cloak-service}
    \vspace{0.085cm}
    \tcp{This contract is constructed by the config of the executor and the enclave. $pk_{\mathcal{E}}, adr_{\mathcal{E}}$ is the public key and address of the enclave $\mathcal{E}$, where $pk_{\mathcal{E}}$ is used for parties building secure channels with the $\mathcal{E}$ and $adr_{\mathcal{E}}$ manages coins deposited by parties and the executor. $adr_{\textit{\textbf{E}}}$ is the address of the executor. For simplicity, we omit access control logic here, but remark it each function.} 
    \vspace{0.085cm}
    \SetKwFunction{FConstructor}{\textit{constructor}}
    \SetKwProg{Fn}{Function}{}{}
    \Fn{\FConstructor{$pk_{\mathcal{E}}, adr_{\mathcal{E}}, adr_{\textbf{\textit{E}}}$}}{
        $pk_{\mathcal{E}}, adr_{\mathcal{E}} \gets pk_{\mathcal{E}}, adr_{\mathcal{E}}$ \tcp{for secure channel} 
        $adr_{\textbf{\textit{E}}} \gets adr_{\textbf{\textit{E}}} $ \\
        $Proposals \gets []$ \\
    }
    \vspace{0.085cm}
    \SetKwFunction{FRegister}{\textit{register}}
    \SetKwProg{Fn}{Function}{}{}
    \Fn{\FRegister{$pk_i$}}{
        \tcp{called by $TX_{pk}$}
        $PartyPKs[msg.sender] \gets pk_i$
    }
    \vspace{0.085cm}
    \SetKwFunction{FDeposit}{\textit{deposit}}
    \SetKwProg{Fn}{Function}{}{}
    \Fn{\FDeposit{$Q$}}{
        \tcp{called by $TX_{col}$ from parties and the executor}
        $Coins[msg.sender] \gets Coins[msg.sender] + Q$
    }
    \SetKwFunction{FPropose}{\textit{propose}}
    \SetKwProg{Fn}{Function}{}{}
    \Fn{\FPropose{$id_p, p', \tau_{com}$}}{
        \tcp{called by $TX_p$ from $\mathcal{E}$ to settle an \acrshort{mpt} proposal} 
        $\FuncSty{require}(Proposals[id_p] = \emptyset)$ \\
        $Proposals[id_p].\{C_f, C_{\mathcal{P}}, \bar{\textit{\textbf{P}}}, C_{x}, q, \tau_{com}, h_{cp}\} \gets p'.\{C_f, C_{\mathcal{P}}, \bar{\textit{\textbf{P}}}, C_{x}, q\}, \tau_{com}, \textbf{\textit{BC}}.\FuncSty{getHeight}()$ \\
        \tcp{deduct collaterals before execution}
        $Coins[adr_{\textbf{\textit{E}}}] \gets Coins[adr_{\textbf{\textit{E}}}] - |\bar{\textit{\textbf{P}}}|*Proposals[id_p].q$ \\
        $\FuncSty{require}(Coins[adr_{\textbf{\textit{E}}}] \geq 0)$ \\
        \textbf{for} $P_i \in \bar{\textit{\textbf{P}}}$ \textbf{do} \\
            \quad$Coins[P_i] \gets Coins[P_i] - Proposals[id_p].q$ \\
            \quad$\FuncSty{require}(Coins[P_i] \geq 0)$ \\
        $Proposals[id_p].st \gets \texttt{SETTLE}$
    }
    \vspace{0.085cm}
    \SetKwFunction{FChallenge}{\textit{challenge}}
    \SetKwProg{Fn}{Function}{}{}
    \Fn{\FChallenge{$id_p, \bar{\textit{\textbf{P}}}_M, data_{cha}$}}{
        \tcp{called by $TX_{cha}$ from $\mathcal{E}$ to challenge specific parties} 
        $\FuncSty{require}(Proposals[id_p].st = \texttt{SETTLE})$ \\
        \textbf{for} $P_i \in \bar{\textit{\textbf{P}}}_M$ \textbf{do} \\
            \quad$Proposals[id_p].\bar{\textit{\textbf{P}}}[P_i].challenge \gets data_{cha} $ \\
    }
    \vspace{0.085cm}
    \SetKwFunction{FResponse}{\textit{response}}
    \SetKwProg{Fn}{Function}{}{}
    \Fn{\FResponse{$id_p, data_{res}$}}{
        \tcp{called by $TX_{res}$ from parties being challenged} 
        $\FuncSty{require}(Proposals[id_p].st = \texttt{SETTLE})$ \\
        $Proposals[id_p].\bar{\textit{\textbf{P}}}[msg.sender].response \gets data_{res} $ \\
    }
    \vspace{0.085cm}
    \SetKwFunction{FPunish}{\textit{punish}}
    \SetKwProg{Fn}{Function}{}{}
    \Fn{\FPunish{$id_p, \bar{\textit{\textbf{P}}}'_M$}}{
        \tcp{called by $TX_{pus}$ from $\mathcal{E}$}
        $\FuncSty{require}(Proposals[id_p].st = \texttt{SETTLE})$ \\
        $refunds = Proposals[id_p].q*(1 + \frac{|\bar{\textit{\textbf{P}}}'_M|}{|Proposals[id_p].\bar{\textit{\textbf{P}}}-\bar{\textit{\textbf{P}}}'_M|+1})$ \\
        \textbf{for} $P_i \in (Proposals[id_p].\bar{\textit{\textbf{P}}}-\bar{\textit{\textbf{P}}}'_M)$ \textbf{do}\\
            \quad$Coins[P_i] \gets Coins[P_i] + refunds$ \\
        $Coins[adr_{\textbf{\textit{E}}}] \gets Coins[adr_{\textbf{\textit{E}}}] + refunds$ \\
        $Proposals[id_p].st \gets \texttt{ABORT}$
    }
    \vspace{0.085cm}
    \SetKwFunction{FMain}{\textit{complete}}
    \SetKwProg{Fn}{Function}{}{}
    \Fn{\FMain{$id_p, proof, C_{s}, C_{s'}, C_{r}, k$}}{
        \tcp{called by $TX_{com}$ from $\mathcal{E}$ to punish misbehaved parties} 
        $\FuncSty{require}(Proposals[id_p].st = \texttt{SETTLE})$ \\
        \textbf{if} $\FuncSty{verify}(proof, C_f, C_{\mathcal{P}}, Proposals[id_p].C_{x}, C_{s}, C_{s'}, C_{r})$ \textbf{then} \\
            \quad$\FuncSty{setNewState}(C_{s'})$ \\
            \quad$Proposals[id_p].\{C_{r}\} \gets \{C_{r}\}$ \\
            \quad$refunds = Proposals[id_p].q$ \\
            \quad\textbf{for} $P_i \in Proposals[id_p].\bar{\textit{\textbf{P}}}$ \textbf{do}\\
                \quad\quad$Coins[P_i] \gets Coins[P_i] + refunds$ \\
            \quad$Coins[adr_{\textbf{\textit{E}}}] \gets Coins[adr_{\textbf{\textit{E}}}] + refunds$ \\
            \quad$Proposals[id_p].st \gets \texttt{COMPLETE}$ \\
    }
    \vspace{0.085cm}
    \SetKwFunction{FSettle}{\textit{timeout}}
    \SetKwProg{Fn}{Function}{}{}
    \Fn{\FSettle{$id_p$}}{
        \tcp{called by $TX_{out}$ from parties} 
        $\FuncSty{require}(\textbf{\textit{BC}.\FuncSty{getHeight}()}>Proposals[id_p].h_{cp} + \tau_{com})$ \\
        $\FuncSty{require}(Proposals[id_p].st = \texttt{SETTLE})$ \\
        $refunds = Proposals[id_p].q * 2$ \\
        \textbf{for} $P_i \in Proposals[id_p].\bar{\textit{\textbf{P}}}$ \textbf{do} \\
            \quad$Coins[P_i] \gets Coins[P_i] + refunds$ \\
        $Proposals[id_p].st \gets \texttt{ABORT}$
    }
    \vspace{0.085cm}
\end{algorithm}

\vspace{-0.15cm}
\subsection{Enclave facility}
The enclave program was implemented as an App of CCF in C++ for reusing CCF's key generation and synchronization functionality. Specifically, CCF is a \acrshort{tee}-based consortium framework. It is easy for us constantly synchronizing a common key pair among all \acrshort{tee} devices in a CCF network. Therefore, we can easily add new \acrshort{tee} device to improve the availability of the executor, without breaking the assumption and requirement of our protocol. Moreover, both CCF and our enclave program are developed based on Openenclave~\cite{openenclave}. Openenclave provides \acrshort{tee}-agnostic API for developing enclave programs, making our implementation easier to adapt to different \acrshort{tee} platforms. 

Our enclave program is presented in Algorithm~\ref{alg:cloak-enclave}. When the executor \textbf{\textit{E}} instantiates a \acrshort{tee} enclave using the enclave program, the enclave becomes $\mathcal{E}$. In more detail, \textbf{\textit{E}} set up $\mathcal{E}$ with a secure parameter $\kappa$ and a checkpoint $b_{cp}$ of blockchain, then, publishes its ${\mathcal{E}}$'s IAS Attestation Report $REP_{ias}$.

To verify/sign transactions as well as building secure channels with parties inside enclave, we port OpenSSL and secp256k1~\cite{secp256k1} to support needed ECDSA. To allow flexible specification of \acrshort{mpt}, \eg, specifying identities who are able/required to join an \acrshort{mpt}, we implement an \textit{policy engine} inside enclave to interpret and enforce JSON-based privacy policy $\mathcal{P}$ of \acrshort{mpt}. The target function of \acrshort{mpt}s are expressed in Solidity 0.8.10~\cite{solc} and we port EVM~\cite{evm} to CCF~\cite{ccf2019}.

\begin{algorithm}[!htbp]
    \footnotesize
    \DontPrintSemicolon
    \LinesNumbered
    \caption{Enclave program ($\mathcal{E}$)}
    \label{alg:cloak-enclave}
    \vspace{0.085cm}
    \tcp{$\mathcal{E}$ is set up with a secure parameter $\kappa$ and a checkpoint $b_{cp}$ of blockchain. $\kappa$ is used for generating an asymmetric key pair $(pk_{\mathcal{E}}, sk_{\mathcal{E}})$ for building secure channels, an blockchain account $(adr_{\mathcal{E}}, key_{\mathcal{E}})$ for managing coins and sending transactions on-chain, and block intervals $\tau_{com}$ for judging \acrshort{mpt} timeout on-chain. For simplicity, we omit the logic of setting up $f, \mathcal{P}, adr_{\mathcal{V}}$}
    \vspace{0.085cm}
    \SetKwFunction{FGsetup}{\textit{setup}}
    \SetKwProg{Fn}{Function}{}{}
    \Fn{\FGsetup{$\kappa, b_{cp}$}}{
        $pk_{\mathcal{E}}, sk_{\mathcal{E}}, adr_{\mathcal{E}}, key_{\mathcal{E}} \gets Gen(1^\kappa)$ \\
        $t_n, t_e, t_d, \tau_{sub}, \tau_{rec}, \tau_{com} \gets Gen(1^\kappa)$ \\
        \Return $pk_{\mathcal{E}}, adr_{\mathcal{E}}$
    }
    \vspace{0.085cm}
    \SetKwFunction{FGenIDp}{\textit{generateIDp}}
    \SetKwProg{Fn}{Function}{}{}
    \Fn{\FGenIDp{$p$}}{
        $id_p \gets Gen(1^\kappa) $ \\
        $t_{cp} \gets \FuncSty{currTime}()$ \\
        $step \gets \texttt{propose}$ \\
        \Return $( id_p, p )$
    }
    \vspace{0.085cm}
    \SetKwFunction{FNpropose}{\textit{propose}}
    \SetKwProg{Fn}{Function}{}{}
    \Fn{\FNpropose{$ACK$}}{
        \textbf{if} $step\neq \texttt{propose}$ \textbf{or} $\FuncSty{SatiPolicy}(ACK, \mathcal{P})\neq1$ \textbf{or} $\FuncSty{currTime}() \geq t_{cp} + t_n $  \textbf{then} abort \\
        $p' \gets ( p.C_f, p.C_{\mathcal{P}}, p.q, \bar{\textit{\textbf{P}}}, ACK.C_{x}) $ \\
        $step \gets \texttt{execute}$ \\
        \Return $TX_p(id_p, p', \tau_{com})$
    }
    \vspace{0.085cm}
    \SetKwFunction{FEexecute}{\textit{execute}}
    \SetKwProg{Fn}{Function}{}{}
    \Fn{\FEexecute{$in, TX_p, PoP_{p}$}}{
        \textbf{if} $step\neq \texttt{execute} $~\textbf{or}~$\FuncSty{veriPoP}(b_{cp}, TX_{p}, PoP_{p})\neq 1$ \textbf{then} abort \\
        $\bar{\textit{\textbf{P}}}_M \gets \emptyset$ \\
        \textbf{for} $P_i$~\textbf{in}~$\bar{\textit{\textbf{P}}}$ \\
            \quad\textbf{if} $in.\{x_i, s_i\} = \emptyset $ \textbf{or}~$\FuncSty{hash}(x_i)\neq PoP_{p}.TX_p.C_{x_i}$ \\
            \quad\quad\textbf{or}~$\FuncSty{hash}(s_i)\neq PoP_{p}.C_{s_i}$ \textbf{then} \\ 
                \quad\quad$\bar{\textit{\textbf{P}}}_M \gets \bar{\textit{\textbf{P}}}_M \cup \{P_i\}$ \\
        \textbf{if} $|\bar{\textit{\textbf{P}}}_M|>0$ \textbf{then} \\
            \quad\Return  $(id_p, \bar{\textit{\textbf{P}}}_M)$   \\
        $out \gets s', r \gets f(s, x)$ \tcp{evaluates $f(x)$ on old states $s$}
        $b_{cp} \gets PoP_{p}.\FuncSty{getLastBlock}() $ \\
        $k \gets Gen(1^\kappa)$ \tcp{generates a symmetric key}
        $step \gets \texttt{complete}$ \\
        \Return $[(id_p, \FuncSty{Enc}_k(out_i))]$ \\
    }
    \vspace{0.085cm}
    \SetKwFunction{FEchallenge}{\textit{challenge}}
    \SetKwProg{Fn}{Function}{}{}
    \Fn{\FEchallenge{$id_p, \bar{\textit{\textbf{P}}}_M$}}{
        \textbf{if} $|\bar{\textit{\textbf{P}}}_M|<0$ \textbf{then} abort \\
        \textbf{if} $step = \texttt{execute}$ \textbf{and} $\FuncSty{currTime}() \leq t_{cp} + t_e $ \textbf{then} \\
            \quad\Return $TX_{cha}(id_p, \bar{\textit{\textbf{P}}}_M, \emptyset)$  \\
        \textbf{elif} $step = \texttt{complete}$ \textbf{and} $\FuncSty{currTime}() \leq t_{cp} + t_d $ \textbf{then} \\
            \quad\Return $TX_{cha}(id_p, \bar{\textit{\textbf{P}}}_M,  [\FuncSty{Enc}_{pk_{i}}(\FuncSty{Enc}_{k}(out_i))])$  \\
        \textbf{else} abort \\
    }
    \vspace{0.085cm}
    \SetKwFunction{FEpunish}{\textit{punish}}
    \SetKwProg{Fn}{Function}{}{}
    \Fn{\FEpunish{$TX_{cha}, TX_{res}, PoP_{res}$}}{
        $\bar{\textit{\textbf{P}}}'_M \gets \emptyset$ \\
        \textbf{for} $P_i \in TX_{cha}.\bar{\textit{\textbf{P}}}_M$ \textbf{do} \\
        \quad\textbf{if} $step = \texttt{execute}$ \textbf{then} \\
            \quad\quad~\textbf{if}~$\FuncSty{veriPoP}(b_{cp}, TX^i_{res}, PoP_{res}, \tau_{sub})\neq 1$ \\ \quad\quad\quad~\textbf{or}~$\FuncSty{hash}(TX^i_{res}.x_i)\neq PoP_{res}.TX_p.C_{x_i}$\\
            \quad\quad\quad~\textbf{or}~$\FuncSty{hash}(TX^i_{res}.s_i))\neq C_{s_i}$ \textbf{then} $\bar{\textit{\textbf{P}}}'_M \gets \bar{\textit{\textbf{P}}}'_M \cup \{P_i\}$ \\
        \quad\textbf{elif}~$step = \texttt{complete}$ \textbf{then} \\
            \quad\quad~\textbf{if}~$\FuncSty{veriPoP}(b_{cp}, TX^i_{res}, PoP_{res}, \tau_{rec})\neq 1$ \\ \quad\quad\quad~\textbf{or}~$TX_{res}.REC_i\neq \FuncSty{hash}(out_i)$ \textbf{then} 
            $\bar{\textit{\textbf{P}}}'_M \gets \bar{\textit{\textbf{P}}}'_M \cup \{P_i\}$ \\
        \quad\textbf{else} abort \\
        \textbf{if} $|\bar{\textit{\textbf{P}}}'_M|>0$ \textbf{then}  \\
            \quad$ step \gets \perp $ \\
            \quad\Return $TX_{pns}(id_p, \bar{\textit{\textbf{P}}}'_M)$ \\
    }
    
    \vspace{0.085cm}
    \SetKwFunction{FDcomplete}{\textit{complete}}
    \SetKwProg{Fn}{Function}{}{}
    \Fn{\FDcomplete{REC}}{
        \textbf{if} $ step\neq \texttt{complete}$ \textbf{or} missed some $REC_i$ \textbf{then} abort \\
        $H_{C_{\mathcal{P}}}, H_{C_f} \gets \FuncSty{hash}(\FuncSty{hash}(\mathcal{P})), \FuncSty{hash}(\FuncSty{hash}(f)) $ \\
        $C_{s'_i}, C_{r_i} \gets \FuncSty{hash}(s'_i), \FuncSty{hash}(r_i)$ \\
        $H_{C_x}, H_{C_s}, H_{C_{s'}}, H_{C_{r}} \gets \FuncSty{hash}(C_x), \FuncSty{hash}(C_s), \FuncSty{hash}(C_{s'}), \FuncSty{hash}(C_r)$ \\
        $proof\gets [H_{C_{\mathcal{P}}}, H_{C_f}, H_{C_x}, H_{C_s}, H_{C_{s'}}, H_{C_{r}}] $ \\
        $step \gets \perp $ \\
        \Return $TX_{com}(id_p, proof, C_{s}, C_{s'}, C_{r}, k)$
    }
    \vspace{0.085cm}
\end{algorithm}

\vspace{-0.15cm}
\subsection{Optimization}\label{sec:optimization}
Instead of reading/writing the whole state of the contract which is adopted in~\cite{Ekiden:2019}. \codename synchronize states with blockchain as need. We pre-specify the states I/O of \acrshort{mpt} in its privacy policy to inform the $\mathcal{E}$ what old states are needed for evaluating the \acrshort{mpt} and what state would be mutated. More details of the policy refer to \cite{ren2021CloakDemo}. Admittedly, reading/writing states according to pre-defiend policy requires that all possible states I/O of an \acrshort{mpt} should be statically recognized before evaluation, so that disallow inputs-depends states I/O logic. We stress that the problem can be totally solved by hooking EVM instructions \code{sload} and \code{sstore} like~\cite{secondstate'20}. We leave this for our future work.  
    
\vspace{-0.15cm}
\section{Security analysis} \label{sec:cloak-security}
    \label{sec:protocol-security}
We informally claim that \codename protocol satisfies five properties: \textbf{\textit{correctness}}, \textbf{\textit{confidentiality}}, \textbf{\textit{public verifiability}}, \textbf{\textit{executor balance security}}, and \textbf{\textit{financial fairness}} in the following theorem. In Appendix~\ref{sec:security-definition} and \ref{sec:security-proof}, we formally define the security properties, state the theorem, and proves our protocol. 

\aptLtoX[graphic=no,type=env]{\begin{theorem}[Informal statement]
\textit{The protocol $\pi_{\codename}$ satisfies \textbf{correctness}, \textbf{confidentiality}, \textbf{public verifiability}, \textbf{executor balance security}, and \textbf{financial fairness}}
\end{theorem}}{
    \begin{informal-theorem}[Informal statement] \textit{The protocol $\pi_{\codename}$ satisfies \textbf{correctness}, \textbf{confidentiality}, \textbf{public verifiability}, \textbf{executor balance security}, and \textbf{financial fairness}}
    \end{informal-theorem}
}


Particularly, as \codename claims to resist a Byzantine adversary, it includes resisting the single-point failure and rollback attack. For the former, Cloak always punishes the executor presenting single-point failure, thinking an honest executor can improve its availability by various schemes, \eg, multiple TEEs with consensus and synchronized keys (as we implemented in Section~\ref{sec:implementation}). For the later, first, $\mathcal{V}$ always rejects \acrshort{mpt}'s outputs from unmatched states, \eg, rollbacked states. Second, the \acrshort{tee} is initiated with a checkpoint block $b_{cp}$, and it always validates the \acrshort{pop} and updates the $b_{cp}$ when read data (\eg, contract states and parameter commitments) on blockchain. Thanks to \acrshort{pop} which ensures that the data to read have been finalized by the consensus, \acrshort{tee} ensures that it always read on-chain data from a monotonically increasing main chain. 


\vspace{-0.15cm}
\section{Evaluation} \label{sec:evaluation}
    \myparagraph{Methodology and setup}
To evaluate the effectiveness of \codename, we propose 3 research questions.
\begin{itemize}[leftmargin=3mm, parsep=0mm, topsep=1mm, partopsep=0mm]
    \item \textbf{Q1}: Does \codename fit real-world needs of publicly verifiable \acrshort{mpt}?
    \item \textbf{Q2}: What's the cost of the deployment and global setup for enabling publicly verifiable \acrshort{mpt} on a blockchain by using \codename?
    \item \textbf{Q3}: What's the cost of evaluating a \acrshort{mpt} by using \codename?
\end{itemize}
To answer \textbf{Q1}, we apply \codename to 5 contracts with 10 different \acrshort{mpt}s. As is shown in Table~\ref{tab:contracts-evaluation}, these contracts vary from both LOC, scenarios, and number of participants. Specifically, their business involves energy, education, and blockchain infrastructure. The involved parties of these 10 \acrshort{mpt} spans from 2 to 11. 

\begin{table}[!htbp]
  \centering
  \vspace{-0.2cm} 
  \caption{\textbf{Contracts with \acrshort{mpt}}. \small{``\#\acrshort{mpt}'' denotes the number of \acrshort{mpt} and  ``\textbf{Scenarios}'' denotes typical business.}}
  \vspace{-0.2cm}
  \label{tab:contracts-evaluation}
  \setlength{\tabcolsep}{0.95mm}{
      \small
      \begin{tabular}{lccc}
        \toprule
        \textbf{Name} & \textbf{\#\acrshort{mpt}} & \textbf{\#LOC} & \textbf{Scenarios} \\
        \midrule
        \code{SupplyChain} & \textbf{1} & 39 & \makecell[l]{An example contract allowing suppliers\\to negotiate and privacy-preservedly bids\\off-chain, and commit the evaluation\\with their new balances on-chain} \\ 
        \cmidrule{4-4} \code{Scores} & \textbf{1} & 95 & \makecell[l]{An example contract allowing students\\to join and get mean scores off-chain\\and commit the evaluation on-chain} \\ 
        \cmidrule{4-4} \code{ERC20Token} & \textbf{3} & 55 & \makecell[l]{An example contract allowing accounts\\to pair and transfer without revealing\\balances off-chain, and commit the\\evaluation with new balances on-chain.} \\
        \cmidrule{4-4} \code{YunDou} & \textbf{3} & 105 & \makecell[l]{A real-world token contract supporting\\ co-managed accounts, in which a suffi-\\-cient number of managers self-selectly\\vote to transfer tokens without\\revealing the votes.}\\
        \cmidrule{4-4} \code{Oracle} & \textbf{2} & 60 & \makecell[l]{A real-world Oracle contract that allows \\parties to negotiate to join then jointly\\and verifiably generate random numbers} \\
        \bottomrule
      \end{tabular}
  }
  \vspace{-0.2cm}
\end{table}

To answer \textbf{Q2} and \textbf{Q3}, We record the cost of time and gas for evaluating each \acrshort{mpt}. We also compare \codename with Fastkitten, the SOTA most related to our work. To obtain a comparative experimental setup, considering that Fastkitten is specific for Bitcoin, we implement its on-chain commitment logic as a Solidity smart contract and commit the party inputs of Fastkitten to achieve the comparable public verifiability with \codename. 

The experiment is based on Ubuntu 18.04 with 32G memory and 2.2GHz Intel(R) Xeon(R) Silver 4114 CPU. Although the gas cost of a specific transaction is deterministic, it also varies from transaction arguments. Therefore, we send each \acrshort{mpt} 3 times with different arguments to get the average result.

\vspace{-0.15cm}
\subsection{\textbf{Deployment and Setup Cost}}
To answer \textbf{Q2}, we discuss the gas cost of deploying the contract program $\mathcal{V}$. The result is shown in Figure~\ref{fig:gas-cost}.

\myparagraph{Gas cost of deployment}
In global initialization phase, \codename costs 4.5M gas to deploy the contract program $\mathcal{V}$ to enable \acrshort{mpt} in existing blockchains. This cost is only paid by \codename service provider for once, thus is mostly irrelevant. 

\myparagraph{Gas cost of global setup}
Each party pays 12.7k gas to $register$ (\code{reg.}) its public key and 4.2k gas to $deposit$ (\code{dep.}) its coins. Therefore, this global once paid gas cost is acceptable. 

\vspace{-0.15cm}
\subsection{\textbf{Transaction Cost}}
\begin{figure}[!htbp]
    \vspace{-0.3cm} 
    \setlength{\abovecaptionskip}{-0.15cm} 
    \setlength{\belowcaptionskip}{-0.2cm} 
    \centering
    \includegraphics[width=8.5cm]{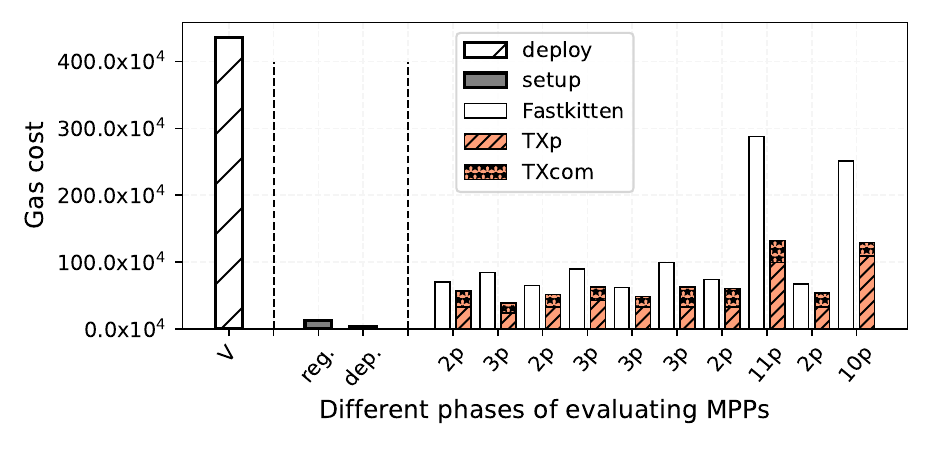}
    \caption{\textbf{The gas cost of \codename}.}
    \label{fig:gas-cost}
\end{figure}

\myparagraph{Gas cost of evaluating \acrshort{mpt}s}
The right part of Figure~\ref{fig:gas-cost} shows the transaction costs of all 10 \acrshort{mpt}s in 5 contracts. In general, \codename reduces gas by 32.4\% compare to Fastkitten, which requires $n+1$ transactions. Specifically, for 4 \acrshort{mpt}s with only 2 parties, \codename cost 0.79-0.82X gas to Fastkitten. However, when the number of parties increases to 10 and 11, the cost of \codename significantly decreases to 0.45-0.46X. Overall, we conclude that \codename evaluates \acrshort{mpt} in not only a securer adversary model but also a lower cost.

\myparagraph{Off-chain latency of evaluating \acrshort{mpt}s} The end-to-end time to evaluate an \acrshort{mpt} is 10 minutes. However, only about 1s is spent on evaluation, with the remainder spent on waiting for the blockchain to generate an \acrshort{pop} of $TX_p$. Precisely, the negotiation phase takes 0.1-0.39s while the execution and distribution phases take 0.26-0.71s. While verifying the $PoP_p$ takes 4-6s, the enclave have to wait 10 minutes for the confirmation of $TX_p$ and the generation of $PoP_p$. We note that the time cost of \acrshort{pop} is common to that of other current \acrshort{tee}-Blockchain systems~\cite{Cavallaro2019Tesseract, Ekiden:2019, FastKitten'19} and is acceptable in permissionless blockchains. More importantly, in widespread quorum-based consensus, the time cost of generating \acrshort{pop} can be reduced to milliseconds~\cite{ccf2019, CONFIDE:SIGMOD20, androulaki2018hyperledger}. Therefore our protocol is ready for use in real-world applications.
    
\vspace{-0.15cm}
\section{Conclusion} \label{sec:conclusion}
    In this paper, we developed a novel framework, \codename, to enable confidential smart contracts with \acrshort{mpt} on existing blockchains. To the best of our knowledge, \codename advances in these aspects. Specifically, \codename is the first to support parties to securely negotiate \acrshort{mpt} proposals off-chain without knowing or communicating with others during the negotiation phase. \codename is the first to achieve public verifiability of an \acrshort{mpt} while considering both on-chain and off-chain inputs/outputs. Moreover, \codename achieves financial fairness under a Byzantine adversary model. Finally, with all the above properties, \codename requires only 2 transactions, far superior to previous work that allows nondeterministic negotiation and financial fairness but requires $O(n)$ transactions. During our evaluation of \codename in both examples and real-world smart contracts, \codename reduces the gas cost by 32.4\%. In conclusion, \codename achieves low-cost and secure \acrshort{mpt}, thereby paving the way for the publicly verifiable and reusable off-chain \acrshort{mpc}s.

\begin{acks}
    This work was supported in part by the National Key R\&D Program of China
(No. 2021YFB2700601); in part by the Finance Science and Technology Project of Hainan Province (No. ZDKJ2020009); in part by the National Natural Science Foundation of China (No. 62163011); in part by the Research Startup Fund of Hainan University under Grant KYQD(ZR)-21071.
\end{acks}

\bibliographystyle{ACM-Reference-Format}
\bibliography{bibliography}

%
\appendix
\label{sec:appendix}

\aptLtoXcmd{%
\section{Symbols and Terminology}
\textbf{EUF-CMA}\quad Existential Unforgeability under CMA.\\
\textbf{HE}\quad Homomorphic Encryption.\\
\textbf{IND-CCA2}\quad Indistinguishability under Adaptive CCA.\\
\textbf{MPC}\quad Multi-Party Computation.\\
\textbf{MPT}\quad Multi-Party Transaction.\\
\textbf{PoP}\quad Proof of Publication.\\
\textbf{TEE}\quad Trusted Execution Environment.\\
\textbf{TTP}\quad Trusted Third Party.\\
\textbf{ZKP}\quad Zero-Knowledge Proof.}{\printnoidxglossary[type=\acronymtype, title={Symbols and Terminology}, nonumberlist]}

\section{Assumption analysis}\label{sec:assumption-rationality}
In this section, we mainly demonstrate why our assumption about \acrshort{tee} is practical and rational and discuss how to fine-tune our protocol to tolerant \acrshort{tee} compromisation.

\subsection{\acrshort{tee} assumption rationality}
Here we elaborate why assuming that the correctness and confidentiality of TEE is practical and rational. Specifically, we assume that \textit{\textbf{E}} has full control over the machine and consequently can execute arbitrary code with root privileges. First, a malicious \textit{\textbf{E}} can exploit memory-corruption vulnerabilities~\cite{Biondo2018codereuse} in enclave code through the API between the host process and the enclave. We assume a common code-reuse defense such as control-flow correctness (CFI)~\cite{abadi2009control, burow2017control}, or fine-grained code randomization~\cite{davi2013gadge} to be in place and active. Then, we consider architectural side-channel attacks, \eg, based on caches~\cite{Brasser2017cacheattack}. These attacks can expose access patterns from SGX enclaves (and therefore our \codename prototype). However, this prompted the community to develop several software mitigations~\cite{gruss2017strong, shih2017t, seo2017sgx} and new hardware-based solutions~\cite{noorman2013sancus, costan2016sanctum}. A more serious Micro-architectural side-channel attacks like Foreshadow~\cite{Bulck2018foreshadow} can extract plaintext data and effectively undermine the attestation process \codename relies on, leaking secrets and enabling the enclave to run a different application than agreed on by the parties; however, the vulnerability enabling Foreshadow was already patched by Intel~\cite{l1tf}. Therefore, since all existing attacks targeting SGX are either patched, function-limited, or having accessible countermeasures, it is still practical to assume that confidentiality and integrity of attested SGX devices hold.

\subsection{Tolerating \acrshort{tee} compromisation}

While we assume that the integrity and confidentiality of $\mathcal{E}$ hold, we stress that it is easy to loose the assumption of $\pi_{\codename}$ to that (i) a factor of \acrshort{tee} devices, or even (ii) all \acrshort{tee}s in some types are compromised. 

For (i), informally, we can instantiate a $\mathcal{E}$ as $3f+1$ \acrshort{tee}s with an Byzantine-tolerant consensus (\eg, PBFT). Meanwhile, those \acrshort{tee}s jointly evaluate any \acrshort{mpt} by \acrshort{mpc} protocols that tolerant $\leq f$ Byzantine nodes, \ie, compromised \acrshort{tee}s here. Then, by regarding messages multi-signed by $\geq 2f+1$ \acrshort{tee}s as messages from a unique $\mathcal{E}$, our assumption holds, as well as all claimed properties of $\pi_{\codename}$.

For (ii), we have two ways, \ie, adapting $\pi_{\codename}$ to another secure \acrshort{tee} product or integrating heterogeneous \acrshort{tee} products. For the first, since \codename’s design is \acrshort{tee} agnostic, it is convenient to adapt a $\pi_{\codename}$ implementation from a type of compromised \acrshort{tee} product to another secure \acrshort{tee} without sacrificing any claimed properties. Specifically, while we instantiate the \acrshort{tee} as SGX, other \codename-compatible and high-application-portability \acrshort{tee} products exist. For example, AMD SEV-SNP~\cite{SEV-SNP} has also supported remote VM attestation. HyperEnclave~\cite{JiaATC22HyperEnclave} presents a cross-platform attestable \acrshort{tee} where SGX programs can run with little/no code changes. Therefore, it is easy to adapt \codename to another securer \acrshort{tee}s to maintain the security properties we claimed. For the second, we can instantiate a $\mathcal{E}$ as a group of heterogeneous \acrshort{tee} products, then follow a similar countermeasure of (i) to achieve (ii). For example, say 4 different \acrshort{tee}s, a SGX, SEV-SNP, HyperEncalve, and Keystone maintain a PBFT consensus and \acrshort{mpc} as in handling (i). Consequently, if adversary corrupt no more than one type of \acrshort{tee} product, the $\mathcal{E}$ holds integrity and confidentiality and $\pi_{\codename}$ holds claimed properties.

\section{Definitions and notations}
In this section, we introduce notations of both our architecture and protocol, and formally define the security properties we claimed in Section~\ref{sec:protocol-security}. 

We refer $\mathcal{S}$ to a set, and denote the $n$-ary Cartesian power of the $\mathcal{S}$ to $\mathcal{S}^n \gets \mathcal{S} \times \mathcal{S} \times \cdots \times \mathcal{S}$. For each vector $\mathbf{s}$, $\mathbf{s} \in \mathcal{S}^n$, we refer the $i$-th coordinate of the vector $\mathbf{s}$ to $\mathbf{s}[i]$. Furthermore, an $n$-by-$m$ matrix of elements from $\mathcal{S}$ is denoted as $\mathbf{S} \in \mathcal{S}^{n \times m}$. The element in the $i$-th row and $j$-column of $\mathbf{S}$ is $\mathbf{S}[i][j]$. The $i$-th row of $\mathbf{S}$ is denoted by $\mathbf{S}[i][\cdot]$ and the $j$-th column of $\mathbf{S}$ is denoted by $\mathbf{S}[\cdot][j]$.

\subsection{Coins and multi-party programs} 
To be simple and chain-agnostic, we define a \emph{coin domain} $\mathcal{D}_{coin}$ as a subset of non-negative rational numbers. The \emph{deposit domain} is denoted by $\mathcal{D}_{dep} \gets \mathcal{D}_{coin}\backslash \{0\}$. Every parties should agree on a \acrshort{mpt} proposal with a deposit vector $\mathbf{q}\in \mathcal{D}^n_{dep}$, which means that every party has to make a deposit. The vector $\mathbf{q}'$ defines the final payout to each party of the contract. It must hold that $\sum_{i\in [n]}\mathbf{q'}[i] \leq \sum_{i\in [n]}\mathbf{q}[i]$. This restrictions guarantees that parties cannot create money by evaluating a program.

For secret input/output of parties, let $\mathcal{D}_{s},
\mathcal{D}_{x}, \mathcal{D}_{s'}, \mathcal{D}_{r}$ refer to the
state domain, parameter domain, new state domain and return value domain respectively. These domains are application-specific and defined by the program. Then a old state vector is denoted by $s \in \mathcal{D}^n_{s} $ and a parameter vector is denoted by $x \in \mathcal{D}^n_{x}$. As both the $s, x$ are parties' input of the protocol, we furthermore denote a input matrix as $\mathbf{in} \in \mathcal{D}^n_{s}\times \mathcal{D}^n_{x}$. Specifically, $\mathbf{in} \gets [s^{\mathsf{T}}, x^\mathsf{T}]$, so that $\mathbf{in}[i][\cdot]$ refers to the input of the party $P_i\in\bar{\bm P}$, $\mathbf{in}[\cdot][0]$ refers to $s$, and $\mathbf{in}[\cdot][1]$ refer to $x$. Similarly, $s' \in \mathcal{D}^n_{s'}$ refers to an new state vector and $r \in \mathcal{D}^n_{r}$ refers to a a return value vector. We denote the output matrix as $\mathbf{ou} \in \mathcal{D}^n_{s'}\times \mathcal{D}^n_{r}$. $\mathbf{ou} \gets [s^{'\mathsf{T}}, r^\mathsf{T}]$, where $\mathbf{ou}[i][\cdot]$ refers to the output of the party $P_i\in\bar{\bm P}$, $\mathbf{ou}[\cdot][0]$ refers to $s'$ and  $\mathbf{ou}[\cdot][1]$ refer to $r$. We model an $n$-party program $f$ as a polynomial time Turing machine. The program $f$ takes a input matrix $\mathbf{in}$  will output a output matrix $\mathbf{ou}$. We denote a program $f$ with specific policy $\mathcal{P}$ as $f_{\mathcal{P}}$. We stress that the policy $\mathcal{P}$ just holds meta-execution data of an \acrshort{mpt}, \eg, the mapping between party identities and inputs, thus is transparent to the execution process. 

For on-chain commitments, we denote the domain of cryptographic commitments as $\mathcal{D}_{cm}$. Then we have $\mathbf{C}_s, \mathbf{C}_x, \mathbf{C}_{s'}, \mathbf{C}_r \in \mathcal{D}^n_{cm}$, where $\mathbf{C}_s, \mathbf{C}_x, \mathbf{C}_{s'}, \mathbf{C}_r$ refer to the old state commitment vector, parameter commitment vector, new state commitment vector and return value commitment vector respectively. Since only the old state commitments are persisted on-chain before an \acrshort{mpt}, we denote $\mathbf{cm}_{\mathbf{in}} \in \mathcal{D}^n_{cm}$ as the input commitment, \ie, $\mathbf{cm}_{\mathbf{in}} \gets \mathbf{C}_s$. 

Finally, we model an $n$-party program $f_{\mathcal{P}}$ as a polynomial time Turing machine. To formally state our security properties, we formally define what is the \emph{correct evaluation of a program} in Algorithm~\ref{alg:eval}. Specifically, the \aptLtoX[graphic=no,graphic=no]{\texttt{eval}}{$\FuncSty{eval}$} takes an $n$-party program $f_{\mathcal{P}}$, a input matrix $\mathbf{in}$ and a input commitment matrix $\mathbf{cm}_{\mathbf{in}}$, and a agreed deposit vector $\mathbf{q}$. The output of the algorithm is the tuple $(s', r, \mathbf{C}_{s'}, \mathbf{C}_{r}, \mathbf{C}_{x}, proof, \mathbf{q'})$, which includes a new state vector $s'$, return value vector $r$, new state commitment vector $\mathbf{C}_{s'}$, return value commitment vector $\mathbf{C}_{r}$, parameter commitment vector $\mathbf{C}_{x}$, a coin distribution vector $\mathbf{q'}$, and the status $st \gets \code{COMPLETE}$ and $proof$ of the evaluation.

\vspace{-0.2cm}
\begin{algorithm}[!htbp]
    \caption{Evaluation function (\FuncSty{eval})}
    \label{alg:eval}
    \LinesNumbered
    \small
    \KwIn{An $n$-party program $f_{\mathcal{P}}$, a deposit vector $\mathbf{q}$, a input matrix $\mathbf{in}$, a old state commitment vector $\mathbf{cm}_{\mathbf{in}}$} 
    \KwOut{A new state vector $s'$, return value vector $r$, new state commitment vector $\mathbf{C}_{s'}$, return value commitment vector $\mathbf{C}_{r}$, parameter commitment vector $\mathbf{C}_{x}$, a \acrshort{mpt} $proof$, and a coin distribution $\mathbf{q'}$} 
 
    \SetKwFunction{Fevaluate}{\FuncSty{eval}}
    \SetKwProg{Fn}{Function}{}{}
    \Fn{\Fevaluate{$f_{\mathcal{P}}, \mathbf{q}, \mathbf{in}, \mathbf{cm}_{\mathbf{in}}$}}{
        \textbf{foreach} $\mathbf{in}.s$ \tcp{foreach $\mathbf{in}[\cdot][0]$}
        \quad \textbf{assert} $\FuncSty{hash}(\mathbf{in}.s[i])=\mathbf{cm}_{\mathbf{in}}[i]$ \tcp{$\FuncSty{hash}(s[i])=\mathbf{C}_{s[i]}$} 
        $ s', r \gets f_{\mathcal{P}}(\mathbf{in}.s, \mathbf{in}.x)$ \\
        $\mathbf{C}_{s'[i]}, \mathbf{C}_{r[i]}, \mathbf{C}_{x[i]} \gets \FuncSty{hash}(s'[i]), \FuncSty{hash}(r[i]), \FuncSty{hash}(\mathbf{in}[i][1])$ \\
        
        $proof\gets [H_{\mathbf{C}_{\mathcal{P}}}, H_{C_f}, H_{C_s}, H_{C_x}, H_{\mathbf{C}_{s'}}, H_{\mathbf{C}_{r}}] $ \\
        $\mathbf{q'} \gets \mathbf{q}$  \\
        \Return $(s', r, \mathbf{C}_{s'}, \mathbf{C}_{r}, \mathbf{C}_{x}, proof, \mathbf{q'})$
    }
\end{algorithm}

\subsection{Protocol execution} 
We consider $n$ parties $\bar{\bm P}$ and a unique executor
${\bm E}$ proceeds \codename protocol $\pi_{\codename}$.
We denote the set including all parties and the executor as $\bar{\bm
P}^+ \gets \bar{\bm P}\cup {\bm E}$.

We assume that all parties $P_i \in \bar{\bm P}$ communicate with the
executor ${\bm E}$ in authenticated channels. According to our
adversary model in Section~\ref{sec:overview}, a protocol is
proceeded in presence of an strong adversary $\mathcal{A}$ who can
arbitrarily corrupt subjects in $\bar{\bm P}^+$. On corrupted
subjects, the $\mathcal{A}$ takes complete control so that
$\mathcal{A}$ can learns and decide the inputs, outputs and also the internal state of the subjects. The input of the protocol execution is an $n$-party program $f_{\mathcal{P}}$, a specified deposit vector $\mathbf{q}$, a input matrix $\mathbf{in}$, and a vector of account balances $\mathbf{Q} \in \mathcal{D}^{n+1}_{dep} \subseteq \mathcal{D}^{n+1}_{coin}$, \ie, the vector of coins pre-deposited to the address $pk_{\mathcal{E}}$ by all subjects in $\bar{\bm P}$. The account domain $\mathcal{D}^n_{dep}$ is defined such that $\forall P_i \in \bar{\bm P}: \mathbf{Q}[i]\geq \mathbf{q}[i]$ and $\mathbf{Q}[i+1] \geq \sum_{P_i\in \bar{\bm P}}\mathbf{q}[i]$. This restriction guarantees that every subject in $\bar{\bm P}^+$ has enough coins for joining an \acrshort{mpt}. We define the execution of the \codename protocol $\pi_{\codename}$ in presence of an adversary $\mathcal{A}$ as
$$\mathbf{ou}, \mathbf{cm}_{\mathbf{ou}}, proof, st, \mathbf{Q}' \gets REAL_{\pi, \mathcal{A}}(\mathbf{Q}, f_{\mathcal{P}}, \mathbf{q}, \mathbf{in}, \mathbf{cm}_{\mathbf{in}})$$

The out commitment matrix is denoted as $\mathbf{cm}_{\mathbf{ou}} \in \mathcal{D}^{n*3}_{cm}$, where $\mathbf{cm}_{\mathbf{ou}} \gets [\mathbf{C}_{s'}^\mathsf{T}, \mathbf{C}_{r}^\mathsf{T}, \mathbf{C}_{x}^\mathsf{T}]$, \ie, $\mathbf{cm}_{\mathbf{ou}}[\cdot][0]$ refers to $\mathbf{C}_{s'}$, $\mathbf{cm}_{\mathbf{ou}}[\cdot][1]$ refers to $\mathbf{C}_r$ and $\mathbf{cm}_{\mathbf{ou}}[\cdot][2]$ refers to $\mathbf{C}_x$. 
The $\mathbf{Q}'\in \mathcal{D}^{n+1}_{coin}$ is the balance vector after the protocol execution. 
The status of the \acrshort{mpt} $st \in \{\emptyset,
\code{COMPLETE}, \code{ABORT}\}$, where $\emptyset$ denotes that the negotiation does not succeeds, the $\code{ABORT}$ means that the negotiation succeeds but the evaluation does not successfully complete. In the case where all subjects in $\bar{\bm P}^+$ are honest, we write the protocol execution as 
$$\mathbf{ou}, \mathbf{cm}_{\mathbf{ou}}, proof, st, \mathbf{Q}' \gets REAL_{\pi}(\mathbf{Q}, f_{\mathcal{P}}, \mathbf{q}, \mathbf{in}, \mathbf{cm}_{\mathbf{in}})$$

\subsection{Security definitions}\label{sec:security-definition}
To better sketch the ability of the $\mathcal{A}$, we denote the set
of honest parties in $\bar{\bm P}^+$  as $\bar{\bm P}^+_{H}$, and the
set of malicious subjects in $\bar{\bm P}^+$ as $\bar{\bm P}^+_{M}$,
\ie, $\bar{\bm P}^+_{M}\gets \bar{\bm P}^+\backslash\bar{\bm
P}^+_{H}$. Similarly, the honest parties in only $\bar{\bm P}$ is
denoted as $\bar{\bm P}_{H}$ and the malicious parties in only
$\bar{\bm P}$ is denoted as $\bar{\bm P}_{M} \gets \bar{\bm
P}\backslash \bar{\bm P}_{H}$.

We first define the basic \emph{correctness} property. Intuitively,
\emph{correctness} states that if all subjects in $\bar{\bm P}^+$
behave honestly, every party $P_i\in \bar{\bm P}$ get correct output and get their collateral back. 

    \begin{definition}[Correctness]
        Protocol $\pi_{\codename}$ run by honest subjects $\bar{\bm
		P}^+$ satisfies the \emph{correctness} property if for every $n$-party program $f_{\mathcal{P}}$, $\mathbf{q}\in \mathcal{D}^n_{dep}$, $s \in \mathcal{D}^n_s$, $x\in \mathcal{D}^n_x$ and $\mathbf{Q}\in\mathcal{D}^{n+1}_{dep}$, the output of the protocol $REAL_{\pi}(\mathbf{Q}, f_{\mathcal{P}}, \mathbf{q}, \mathbf{in}, \mathbf{cm}_{\mathbf{in}})$ is that $\forall P_i \in \bar{\bm P}$~:
        $$Pr[\mathbf{ou}[i][\cdot] \gets [s'[i], r[i]]~\text{and}~\mathbf{Q}'[i]\geq \mathbf{Q}[i]]=1$$
    \end{definition}
The $(s', r, \mathbf{C}_{s'}, \mathbf{C}_{r}, \mathbf{C}_{x}, proof, \mathbf{q'}) \gets$ \aptLtoX[graphic=no,type=env]{\texttt{eval}}{$\FuncSty{eval}$}$(f_{\mathcal{P}}, \mathbf{q}, \mathbf{in}, \mathbf{cm}_{\mathbf{in}})$.

Next, we define the \emph{confidentiality}. 
    \begin{definition}[Confidentiality]
        Protocol $\pi_{\codename}$ run by subjects $\bar{\bm P}^+$
		satisfies the \emph{confidentiality} property if for every
		$n$-party program $f_{\mathcal{P}}$, for every adversary
		$\mathcal{A}$ corrupting parties from $\bar{\bm P}^+$, for
		every $\mathbf{q}\in \mathcal{D}^n_{coin}$, $s \in
		\mathcal{D}^n_s$, $x\in \mathcal{D}^n_x$ and $\mathbf{Q}\in\mathcal{D}^{n+1}_{dep}$,  the protocol $REAL_{\pi, \mathcal{A}}(\mathbf{Q}, f_{\mathcal{P}}, \mathbf{q}, \mathbf{in}, \mathbf{cm}_{\mathbf{in}})$ is such that: $\forall s^{'}_*\in \mathcal{D}_{s'}, r_*\in \mathcal{D}_r$, it holds that $\forall \mathcal{A}$ corrupting parties in $\bar{\bm P}_{M} \cup \{{\bm E}\}$ where $\bar{\bm P}_{M} \subsetneqq \bar{\bm P}$
        $$Pr[\mathbf{ou}[j][\cdot] = [s'[j], r[j]]\mid {P_j \in
		\bar{\bm P}_{H}}]=Pr[\mathbf{ou}[j][\cdot] = [s^{'}_*, r_*]]$$
    \end{definition}
The $(s', r, \mathbf{C}_{s'}, \mathbf{C}_{r}, \mathbf{C}_{x}, proof, \mathbf{q'}) \gets$ \aptLtoX[graphic=no,type=env]{\texttt{eval}}{$\FuncSty{eval}$}$(f_{\mathcal{P}}, \mathbf{q}, \mathbf{in}, \mathbf{cm}_{\mathbf{in}})$.


satisfies the \emph{State availability} property if for every
$n$-party program $f_{\mathcal{P}}$, for every adversary $\mathcal{A}$ corrupting parties from $\bar{\bm P}^+$, for every $\mathbf{q}\in \mathcal{D}^n_{coin}$, $s \in \mathcal{D}^n_s$, $x\in \mathcal{D}^n_x$ and $\mathbf{Q}\in\mathcal{D}^{n+1}_{dep}$, the output of the protocol $REAL_{\pi, \mathcal{A}}(\mathbf{Q}, f_{\mathcal{P}}, \mathbf{q}, \mathbf{in}, \mathbf{cm}_{\mathbf{in}})$ is such that $\forall P_i \in \bar{\bm P}_{H}$~:

Next, we formally define the \emph{public verifiability}.
    \begin{definition}[Public verifiability]
        Protocol $\pi_{\codename}$ run by subjects $\bar{\bm P}^+$
		satisfies the \emph{Public verifiability} property if for every $n$-party program $f_{\mathcal{P}}$, for every adversary $\mathcal{A}$ corrupting parties from $\bar{\bm P}^+$, for every $\mathbf{q}\in \mathcal{D}^n_{coin}$, $s \in \mathcal{D}^n_s$, $x\in \mathcal{D}^n_x$ and $\mathbf{Q}\in\mathcal{D}^{n+1}_{dep}$, the output of the protocol $REAL_{\pi, \mathcal{A}}(\mathbf{Q}, f_{\mathcal{P}}, \mathbf{q}, \mathbf{in}, \mathbf{cm}_{\mathbf{in}})$ is such that both of the following must be true:
        
        \begin{math}
            \begin{aligned}
                & \bullet~\forall~ (s', r, \mathbf{C}_{s'}, \mathbf{C}_{r}, \mathbf{C}_{x}, proof, \mathbf{q'}) \gets \aptLtoX[graphic=no,type=env]{\texttt{eval}}{\FuncSty{eval}}(f_{\mathcal{P}}, \mathbf{q}, \mathbf{in}, \mathbf{cm}_{\mathbf{in}}): \\  
                & \qquad Pr[\aptLtoX[graphic=no,type=env]{\texttt{verify}}{\FuncSty{verify}})(proof, \mathbf{cm}_{\mathbf{in}}, \mathbf{cm}_{\mathbf{ou}}, H_{f}, H_{\mathcal{P}})=1]=1  \\
                & \bullet~\forall~ (s', r, \mathbf{C}_{s'}, \mathbf{C}_{r}, \mathbf{C}_{x}, proof, \mathbf{q'}) \gets \aptLtoX[graphic=no,type=env]{\texttt{eval}}{\FuncSty{eval}}(f_{\mathcal{P}}, \mathbf{q}, \mathbf{in}, \mathbf{cm}_{\mathbf{in}}): \\
                & \qquad Pr[\aptLtoX[graphic=no,type=env]{\texttt{verify}}{\FuncSty{verify}}(proof, \mathbf{cm}_{\mathbf{in}}, \mathbf{cm}_{\mathbf{ou}}, H_{f}, H_{\mathcal{P}})=1]=0  \\
            \end{aligned}
        \end{math}
    \end{definition}

Then we define the security property \emph{executor balance security} which means the executor cannot lose money if it behaves honestly.
    \begin{definition}[Executor balance security]
        Protocol $\pi_{\codename}$ run by subjects $\bar{\bm P}^+$
		satisfies the \emph{executor balance security} property if for every $n$-party program $f_{\mathcal{P}}$, for every adversary $\mathcal{A}$ corrupting only parties from $\bar{\bm P}$ (the executor is honest), for every $\mathbf{q}\in \mathcal{D}^n_{coin}$, $s \in \mathcal{D}^n_s$, $x\in \mathcal{D}^n_x$ and $\mathbf{Q}\in\mathcal{D}^{n+1}_{dep}$, the output of the protocol $REAL_{\pi, \mathcal{A}}(\mathbf{Q}, f_{\mathcal{P}}, \mathbf{q}, \mathbf{in}, \mathbf{cm}_{\mathbf{in}})$ is such that:
        $$Pr[\mathbf{Q}'[n+1]\geq \mathbf{Q}[n+1]]=1$$
    \end{definition}

Finally, we define \emph{financial fairness} which in high level
states that if at least one party $P_i\in \bar{\bm P}$ is honest, then must cause one of the following two events: (i) the protocol correctly evaluates the program and delivers the outputs; (ii) all honest parties output $\code{ABORT}$, stay financially neutral and at least one corrupt party must have been punished on-chain. 
    \begin{definition}[Financial fairness]
        \label{def:fairness}
        Protocol $\pi_{\codename}$ run by subjects $\bar{\bm P}^+$
		satisfies the \emph{financial fairness} property if for every
		$n$-party program $f_{\mathcal{P}}$, for every adversary $\mathcal{A}$ corrupting parties from $\bar{\bm P}^+_{M} \subsetneqq \bar{\bm P}^+$, for every $\mathbf{q}\in \mathcal{D}^n_{coin}$, $s \in \mathcal{D}^n_s$, $r\in \mathcal{D}^n_r$ and $\mathbf{Q}\in\mathcal{D}^{n+1}_{dep}$, the output of the protocol $REAL_{\pi, \mathcal{A}}(\mathbf{Q}, f_{\mathcal{P}}, \mathbf{q}, \mathbf{in}, \mathbf{cm}_{\mathbf{in}})$ is such that one of the following statements must be true:
        
        \begin{math} 
            \begin{aligned}
                & \text{(i)}~st=\emptyset,~\forall P_i\in \bar{\bm P}_{H}: \mathbf{Q}'[i]\geq \mathbf{Q}[i]\\
                & \text{(ii)}~st=\code{ABORT},~\forall P_i\in
				\bar{\bm P}_{H}:~\mathbf{Q}'[i]\geq \mathbf{Q}[i]~\text{and} \\
                & \qquad {\sum_{j\in \bar{\bm P}^+_{M}}
				\mathbf{Q}'[j] < \sum_{j\in \bar{\bm P}^+_{M}} \mathbf{Q}[j]}\\
                & \text{(iii)}~st=\code{COMPLETE},~\forall P_i\in
				\bar{\bm P}_{H}: \mathbf{Q}'[i]\geq \mathbf{Q}[i] -\mathbf{q}[i] + \mathbf{q'}[i] \\
            \end{aligned}
        \end{math}
    \end{definition}
\section{Security proof of Cloak protocol}\label{sec:security-proof}
We have informally explained the main theorem of $\pi_{\codename}$ in Section~\ref{sec:protocol-security}. Here we formally state and prove the theorem.

\begin{theorem}[Formal statement]
    Assume a \acrshort{cma} secure signature scheme, a \acrshort{cca2} encryption scheme, a hash function that is collision-resistant, preimage and second-preimage resistant. a \acrlong{tee} emulating the TEE ideal functionality and a blockchain emulating the blockchain ideal functionality, the \codename protocol $\pi_{\codename}$ satisfies \textbf{correctness}, \textbf{confidentiality}, \textbf{public verifiability}, \textbf{executor balance security}, and \textbf{financial fairness} properties.
\end{theorem}

\subsection{Proof of correctness}
As is defined by \emph{correctness}, we consider the scenario when
all subjects in $\bar{\bm P}^+$ are honest. To evaluate an
\acrshort{mpt}, $\pi_{\codename}$ starts from \emph{Negotiation
phase}. Each party in $\bar{\bm P}$ independently interacts with both
blockchain and the executor \textit{\textbf{E}} to agree to an
\acrshort{mpt} proposal. Once the proposal is confirmed by $TX_{p}$
on the blockchain, the collateral of each party $P_i\in \bar{\bm P}$ is also deducted, which means the coin balance of each party $P_i\in \bar{\bm P}$ becomes $\mathbf{Q}[i]-\mathbf{q}[i]$. Next, the protocol proceeds to the \emph{Execution phase}. In this phase, for all $P_i\in {\bm P}$ the following holds: (1) $P_i$ sends input $in_i\gets(s[i], x[i])$ to the executor \textit{\textbf{E}}. The \textit{\textbf{E}} (2) confirms that the input is correctly signed, then loads the input vector $in$ with the blockchain view into the enclave $\mathcal{E}$. The $\mathcal{E}$ will again verify the signatures of parties and (3) additionally verify that the input $(s[i], x[i])$ match the confirmed input commitments on the blockchain. Then $\mathcal{E}$ evaluate the program $f_{\mathcal{P}}$ as $out \gets (s', r) \gets f_{\mathcal{P}}(in.s, in.x)$. Finally, the protocol moves to \emph{Distribution phase}. The $\mathcal{E}$ first generate a symmetric key $k$ and deliver the ciphertext of the output to each parties $Enc_{k}(out[i])$. When $\mathcal{E}$ ensures that all parties have received their corresponding output ciphertext by receiving parties' receipts, the $\mathcal{E}$ outputs transaction $TX_{com}(id_p, proof, \mathbf{C}_s, \mathbf{C}_{s'}, \mathbf{C}_r, k)$, which refunds $\mathbf{q'}[i]$ to party $P_i$ and release $k$ publicly. Hence, since $\mathbf{q'}=\mathbf{q}$, for every $P_i\in \bar{\bm P}$ it holds that $\mathbf{Q}'[i]\gets \mathbf{Q}[i]-\mathbf{q}[i]+\mathbf{q'}[i]\geq \mathbf{Q}[i]$.
\subsection{Proof of public verifiability}
Recall that $H_{*}$ refers to \aptLtoX[graphic=no,type=env]{$\texttt{hash}$(*)$}{$\FuncSty{hash}(*)$}, for a specific \acrshort{mpt} evaluation, $proof$ is $[H_{\mathbf{C}_{\mathcal{P}}}, H_{C_f}, H_{C_s}, H_{C_x}, H_{\mathbf{C}_{s'}}, H_{\mathbf{C}_{r}}]$ signed by $\mathcal{E}$. Since we assume the integrity of $\mathcal{E}$ is guaranteed, the $proof$ is therefore correctly computed inside $\mathcal{E}$ and signed by $pk_{\mathcal{E}}$. Since we assume that the correctness of hash function and $\pi_{\codename}$, the correctness of $proof$ holds. Furthermore, as the signature scheme is \acrshort{cma}, the signature of $TX_{com}$ is unforgable, as well as the signature of $proof$. Therefore, the \emph{public verifiability} holds.


\subsection{Proof of executor balance security}
We distinguish the following cases when the executor is honest.

\emph{(i):} If the \emph{negotiation phase} failed, it means that
either parties in $\bar{\bm P}$ do not successfully fit the settlement condition of the negotiation so that none $TX_{p}$ is released, or the release $TX_p$ failed in being confirmed on the blockchain. In both scenarios, the collateral of the executor for that \acrshort{mpt} will not be deducted on blockchain.

\emph{(ii):} If the parties in $\bar{\bm P}$ agree on an
\acrshort{mpt} proposal during the \emph{Negotiation phase}, it means
the proposal is successfully confirmed on the blockchain so that the collateral of both all parties in $\bar{\bm P}$ and the executor is successfully deducted. Then, if at least one party does not provide correct signed input in \emph{Execution phase} even after the \textit{\textbf{challenge-response submission}} case, then the enclave will output the transaction $TX_{pns}(id_p, \bar{\bm P}'_{M}, q)$ that returns the deposit back to the executor.

\emph{(iii):} Similar to \emph{(ii)}, if both the \emph{Negotiation
phase} and the \emph{Execution phase} successfully completes, the enclave then deliver the ciphertext to all parties. Next, if at least one party does not provide correctly the signed receipt, even after the \textit{\textbf{challenge-response delivery}} case, then the enclave will also output the transaction $TX_{pns}(id_p, \bar{\bm P}'_{M}, q)$ to return the collateral back to the executor.

\emph{(iv)} If the \emph{Negotiation phase}, \emph{Execution phase}, and delivering the ciphertext of outputs  successfully completes, the enclave outputs the transaction $TX_{com}(id_p, proof, \mathbf{C}_{s}, \mathbf{C}_{s'}, \mathbf{C}_r, k)$ that returns the collateral of the \acrshort{mpt} to the executor.

It remains to discuss whether the transactions in cases \emph{(ii)} and \emph{(iii)} are valid when posted to the blockchain by the executor, \ie, $st=\code{SETTLE}$ when these transactions are posted.
The only transaction that modifies $st$ on-chain is $TX_{out}$, which is posted by parties and only accepted after the $\tau_{com}$-th block after $h_{cp}$, where $h_{cp}$ is the height of block with $TX_p$. Let $\delta$ upper bound the block number from a transaction is published to transaction pool to it is included in a block, $\gamma$ upper bound the block time from a transaction is included in a block to it is confirmed, $\lambda$ upper bound the block number for executing the off-chain program, $\epsilon$ upper bound the block number waiting for collecting party inputs and receipts. Then, we set $\tau_{com} \geq 5(\delta + \gamma)+ \lambda + 2\epsilon$. Specifically, starting from the block height $h_{cp}$ where $TX_p$ is included on chain, confirming $TX_p$ costs $\gamma$ blocks first. Because waiting for party inputs costs $\epsilon$, we set $t_e\gets \gamma + \epsilon$. Next, a possible challenge-response submission stage needs to publish and confirm two transactions, which costs $2(\delta + \gamma)$ blocks. Therefore, we set $\tau_{sub}\gets t_e + 2(\delta + \gamma)$. After that, executing the program in enclave costs $\lambda$ blocks and waiting for receipts costs $\epsilon$, so we have $t_d\gets \tau_{sub} + \lambda + \epsilon$. As a possible challenge-response delivery stage costs $2(\delta + \gamma)$, we set $\tau_{rec}\gets \tau_{sup} + \lambda + \epsilon + 2(\delta + \gamma)$. Finally, since publishing $TX_{com}$ and including it in a block costs $\delta$ blocks, we set $\tau_{com}\gets \tau_{rec} + \delta$. In conclusion, the honest executor always has time to publish $TX_{pns}$ or $TX_{com}$ with specified $\tau_{com}$.

\subsection{Proof of financial fairness}
We prove the financial fairness of $\pi_{\codename}$ phases by
phases. First, we consider the \emph{Negotiation phase}. Briefly, we will show that if the \emph{Negotiation phase} does not complete successfully then all honest parties in $\bar{\bm P}_{H}$ will not go into $\code{SETTLE}$ and stay financially neutral.
\vspace{-0.1cm}
\begin{lemma}
        If there exist an honest party $P_i$ stay at $st=\emptyset$, then the statement (i) of the financial fairness property holds.
\end{lemma}
\vspace{-0.2cm}
\emph{Proof:}
According to $\pi_{\codename}$, there are only one case when an honest party $P_i$ will stay at $\emptyset$: 
\begin{itemize}[leftmargin=5mm, parsep=0.5mm, topsep=0.5mm, partopsep=0.5mm]
    \item No $TX_p$ is confirmed on the blockchain. 
\end{itemize}
Specifically, this scenario happens in following reasons: Parties in
$\bar{\bm P}$ either fail to agree on an \acrshort{mpt} proposal,
preventing the enclave from constructing and releasing the $TX_p$, or the $TX_p$ is released but fails to be confirmed on blockchain for reasons like that at least one subject in $\bar{\bm P}^+$ does not have enough coins to be deducted as the collateral for the \acrshort{mpt}. However, no matter what reasons fail the confirmation of $TX_p$, all parties and the executor in $\pi_{\codename}$ only identify the $st$ of an \acrshort{mpt} by reading its confirmed status from the blockchain. Since we have assumed that the blockchain has ideal consistency and availability, all honest parties can access the consistent blockchain view. Therefore, if the $TX_p$ is successfully confirmed on-chain which means that parties' collateral has been successfully deducted, honest parties will immediately identify the \acrshort{mpt} as $\code{SETTLE}$. In other words, if at least an honest party stay at $\emptyset$, the $TX_p$ must be not successfully confirmed so that none of parties' collateral is deducted, \ie, $\mathbf{Q}'[i]\geq \mathbf{Q}[i]$.

Next, we show that the financial fairness also holds even if \acrshort{mpt} failed by $\code{ABORT}$ after an successful \emph{Negotiation phase}.
\begin{lemma}
    If there exist an honest party $P_i$ such that $st = \code{ABORT}$, then the statement (ii) of the financial fairness property holds.
\end{lemma}

\emph{Proof:} Three cases exists when an honest party $P_i$ outputs $\code{ABORT}$:
\begin{itemize}[leftmargin=5mm, parsep=0.5mm, topsep=0.5mm, partopsep=0.5mm]
    \item (i) Before the $\tau_{com}$-th block succeeding to the
	block confirming the $TX_p$, The $TX_{pns}(id_p, \bar{\bm P}'_M, p)$ is published on the blockchain after a \textit{challenge-response submission} stage.
    \item (ii) Before the $\tau_{com}$-th block succeeding to the
	block confirming the $TX_p$, The $TX_{pns}(id_p, \bar{\bm P}'_M, p)$ is published on the blockchain after a \textit{challenge-response delivery} stage.
    \item (iii) After the $\tau_{com}$-th block succeeding to the block confirming the $TX_p$, The $TX_{out}(id_p)$ is published on the blockchain.
\end{itemize}

We first consider the case (i) where malicious parties do not provide
inputs after the negotiation succeeded. According to
Algorithm~\ref{alg:cloak-enclave}, the enclave $\mathcal{E}$ release
a transaction $TX_{pns}(id_p, \bar{\bm P}'_{M}, q)$ if and only if
the executor calls the $\mathcal{E}.punish$ with a blockchain view
which shows that parties in the non-empty set $\bar{\bm P}'_{M}$ did
not provide their inputs even though they were challenged. By the definition of $TX_{pns}$, all honest parties, \ie, parties not in $\bar{\bm P}'_{M}$, and the executor will get their collateral back and the parties in $\bar{\bm P}'_{M}$ get nothing. In other word, for $\forall P_i \in \bar{\bm P}'_{M}$ it holds that $\mathbf{Q}'[i] =\mathbf{Q}[i] - \mathbf{q}[i]$. Since $\mathbf{q}[i]>0$ and $\bar{bm P}'_{M}\neq \emptyset$, at least one malicious party lost coins. Therefore, the inequality (ii) in Definition~\ref{def:fairness} holds.

We then consider the case (ii) where malicious parties do not
response receipts when they received the ciphertext of outputs. According to the definition of Algorithm~\ref{alg:cloak-enclave}, again, the $\mathcal{E}$ output a transaction $TX_{pns}$ if and only if the executor proves that the ciphertext of outputs has been publicly sent to parties as challenges on the blockchain but the parties being challenged did not response with their receipts. Similar to the case (i), for $\forall P_i \in \bar{\bm P}'_{M}$ it also holds that $\mathbf{Q}'[i] = \mathbf{Q}[i] - \mathbf{q}[i]$, \ie, the inequality (ii) in Definition~\ref{def:fairness} holds. 

Finally we consider the case (iii) where indicates a malicious
executor. In this case, the timeout transaction $TX_{out}$ is posted
on the blockchain which mean that every $P_i\in \bar{\bm P}$ gets $\mathbf{q}[i]$ coins back, \ie, $\mathbf{Q}'[i] = \mathbf{Q}[i]$, and the executor loses all its collateral for the evaluation, \ie, $\mathbf{Q}'[n+1] = \mathbf{Q}[n+1] - \sum_{P_i\in \bar{\bm P}}\mathbf{q}[i]$. Because the malicious executor lost collateral and no other malicious party earned any collateral, the inequality (ii) in Definition~\ref{def:fairness} holds. 

    \begin{lemma}
        If there exist an honest party $P_i$ such that $st \notin \code{ABORT}$, then the statement (iii) of the financial fairness property holds.
    \end{lemma}

\emph{Proof:} According to Algorithm~\ref{alg:cloak-service} , the
protocol outputs  $\mathbf{ou}[i] \notin \code{ABORT}$ if and only if
a transaction $TX_{com}(id_p, proof, \mathbf{C}_s, \mathbf{C}_{s'}, \mathbf{C}_r, k)$ is posted on the blockchain before the $h_{cp}+\tau_{com}$-th block. Furthermore, by definition of enclave program in Algorithm~\ref{alg:cloak-enclave}, the enclave $\mathcal{E}$ releases the $TX_{com}$ if and only if all parties have received the ciphertext of their outputs before the $h_{cp}+\tau_{com}$-th block. Since the unique confirmed $TX_{com}$ publishes the $k$ to all parties, each party $P_i \in \bar{\bm P}$ can decrypt the output ciphertext to get $\mathbf{ou}[i]$. Besides, as the $TX_{com}$ also refund the collateral $\mathbf{q'}[i]$ of each party $P_i$ back, we know $\forall P_i \in \bar{\bm P}$ the $\mathbf{Q}'[i] = \mathbf{Q}[i] - \mathbf{q}[i] + \mathbf{q'}[i]$ holds. Therefore, the inequality (i) in Definition~\ref{def:fairness} holds. 


\end{document}